\documentclass[a4paper,11pt]{article}
\usepackage{jheppub} % for details on the use of the package, please see the JINST-author-manual
\newcommand\beq{ \begin{eqnarray} }
\newcommand\eeq{ \end{eqnarray} }

\usepackage{amsmath}
\usepackage{amssymb}
\usepackage{mathrsfs}
\usepackage{graphicx}
\usepackage{bbold}
\usepackage{braket}
\usepackage{bm,bbm}
\usepackage{physics}
\usepackage{tikz}

\arxivnumber{1234.56789} % if you have one

\title{\boldmath Lattice results for the equation of state in dense QCD-like theories}

\author{Etsuko Itou}
\affiliation{Yukawa Institute for Theoretical Physics, Kyoto University, 
Kitashirakawa Oiwakecho, Sakyo-ku, Kyoto 606-8502 Japan\\}
\affiliation{RIKEN Center for Interdisciplinary Theoretical and Mathematical Sciences (iTHEMS), RIKEN,
2-1 Hirosawa, Wako, Saitama 351-0198 Japan}

\emailAdd{itou@yukawa.kyoto-u.ac.jp}

\abstract{
We review recent progress in Monte Carlo simulations of dense two-color QCD (QC$_2$D), focusing on the phase diagram, the equation of state, and the sound velocity in the low-temperature regime. In three-color QCD at finite density, especially at low temperatures, the notorious sign problem makes lattice Monte Carlo simulations intractable. In contrast, QC$_2$D is free from this issue due to the pseudoreality of the quark representation. Recent independent lattice studies have revealed unexpected phenomena through first-principles calculations of the phase structure and thermodynamics.
A particularly notable finding is that the sound velocity exceeds the so-called conformal (holography) bound, \( c_s^2/c^2 \le 1/3 \), which had not been observed in QCD-like theories at finite temperature. In this review, we focus primarily on results from a series of works by our group~\cite{Iida:2019rah, Iida:2020emi,Iida:2022hyy, Ishiguro:2021yxr, Iida:2024irv}, along with related studies in dense QC$_2$D and three-color QCD with isospin chemical potential. We discuss the possibility and physical implications of conformal bound violation even for three-color dense QCD, together with insights from effective model analyses and recent observations of neutron stars. 
}

\begin{document}
\maketitle
\flushbottom

\section{Introduction}
Over the past four decades, lattice QCD has achieved remarkable success as a first-principles approach to nonperturbative dynamics in strong interactions. One of the most notable achievements is the quantitative reproduction of the hadron spectrum: theoretical predictions agree with experimental data at the level of a few percent (see e.g.\ Ref.~\cite{Aoyama:2024cko} and references therein). At finite temperature, lattice QCD has similarly excelled, mapping out the chiral phase transition with high precision and determining the equation of state (EoS) across the crossover region~\cite{Borsanyi:2013bia, HotQCD:2014kol}.
These successes are accomplished with only a few input parameters, namely the gauge coupling and bare quark masses, in the QCD action. 
On the methodological side, the development of the hybrid Monte Carlo (HMC) algorithm has enabled numerically exact simulations in the sense that the algorithm satisfies detailed balance and ergodicity.  In this way, lattice QCD provides a highly successful, gauge-invariant nonperturbative framework both at zero temperature and in the finite-temperature regime.

The situation changes dramatically once a quark chemical potential ($\mu$) is introduced into the QCD action, leading to what is known as dense QCD. The Euclidean action becomes complex, leading to the notorious sign problem.  In 2005, it was shown by M.~Toryer and U.-J.~Wiese that the sign problem is, in general, NP-hard~\cite{Troyer:2004ge}~\footnote{“NP-hard” (nondeterministic polynomial-time hard) is a technical term from computational complexity theory. It indicates that the problem is at least as hard as the most difficult problems whose solutions can be verified in polynomial time.}. Thus, the calculation complexity may grow more than polynomially, in the worst case, exponentially, as the thermodynamic limit is approached.  Despite various attempts to circumvent this issue, reliable simulations at low temperature and high density remain extremely difficult (see, e.g., Refs.~\cite{Aarts:2015kea, Nagata:2021ugx}).

On the other hand, certain QCD-like theories that modify or extend dense QCD are known to be free from the sign problem. These include two-color QCD (QC$_2$D), QCD with isospin chemical potential, and QCD with an imaginary chemical potential ($\mu=i\mu_I$). Although in the imaginary chemical potential case, the value of the chemical potential has an upper limit due to the periodicity of the partition function in $\mu$, the former two systems do not have such a limitation. Indeed, they have attracted growing attention in recent years, especially due to the emergence of novel features in the EoS, such as violations of the conformal (holography) bound on the speed of sound($c_{\rm s}$)~\cite{Hohler:2009tv, Cherman:2009tw}; 
\beq
c_{\rm s}^2/c^2 \le 1/3.
\eeq
In this review, we focus on recent developments, primarily over the past decade, in particular dense QC$_2$D as a representative theory for exploring dense QCD matter.

In QC$_2$D, the sign problem is indeed absent if we consider an even-flavor system.
The fundamental representation, namely quark, of the SU($2$) gauge group takes a (pseudo)real representation~\cite{Nakamura:1984uz, Muroya:2002ry, Muroya:2003qs}. It causes the determinant of the Dirac operator, even in the finite quark chemical potential regime, to be a positive-real or negative-real.
Thus, in the SU(2) gauge theory coupled to an even number of flavors, the determinant always becomes real and positive, and there is no sign problem in the HMC algorithm.

To perform lattice Monte Carlo simulations of QCD(-like) theories, the sign problem is not the only problem, particularly at low temperature and high density.
One must also be careful with the so-called (early) onset problem~\cite{Cohen:2003kd, Muroya:2003qs}. 
The origin of this onset is dynamical pair creation and pair annihilation of the lightest hadrons. Thus, it occurs roughly half the mass of the lightest pseudoscalar(PS) meson, $\mu \gtrsim m_{\rm PS}/2$. 
It is also known as numerical instability; the eigenvalue distribution of the Dirac operator spreads over the complex plane, resulting in some eigenvalues close to the origin that are difficult to compute~\cite{Muroya:2000qp}.
In the case of QC$_2$D, according to the mean-field chiral perturbation theory (ChPT)~\cite{Kogut:2000ek}, the onset scale is predicted to coincide with the superfluid phase transition point where the $U(1)_B$ symmetry is spontaneously broken.
Using this fact, numerical methods have been proposed to avoid the numerical instability problem~\cite {Hands:1999md, Kogut:2001na, Skullerud:2003yc}. Thus, adding the explicit breaking term of the $U(1)_B$ symmetry solves the problem.

%%%%%%%%%%%%%%%%%%%%%%%%%%%%%%%%%%%%%%%%%%%%%%%%%%%%%%%%%%
\begin{figure}[h]
\centering
\includegraphics[width=1.0 \textwidth]{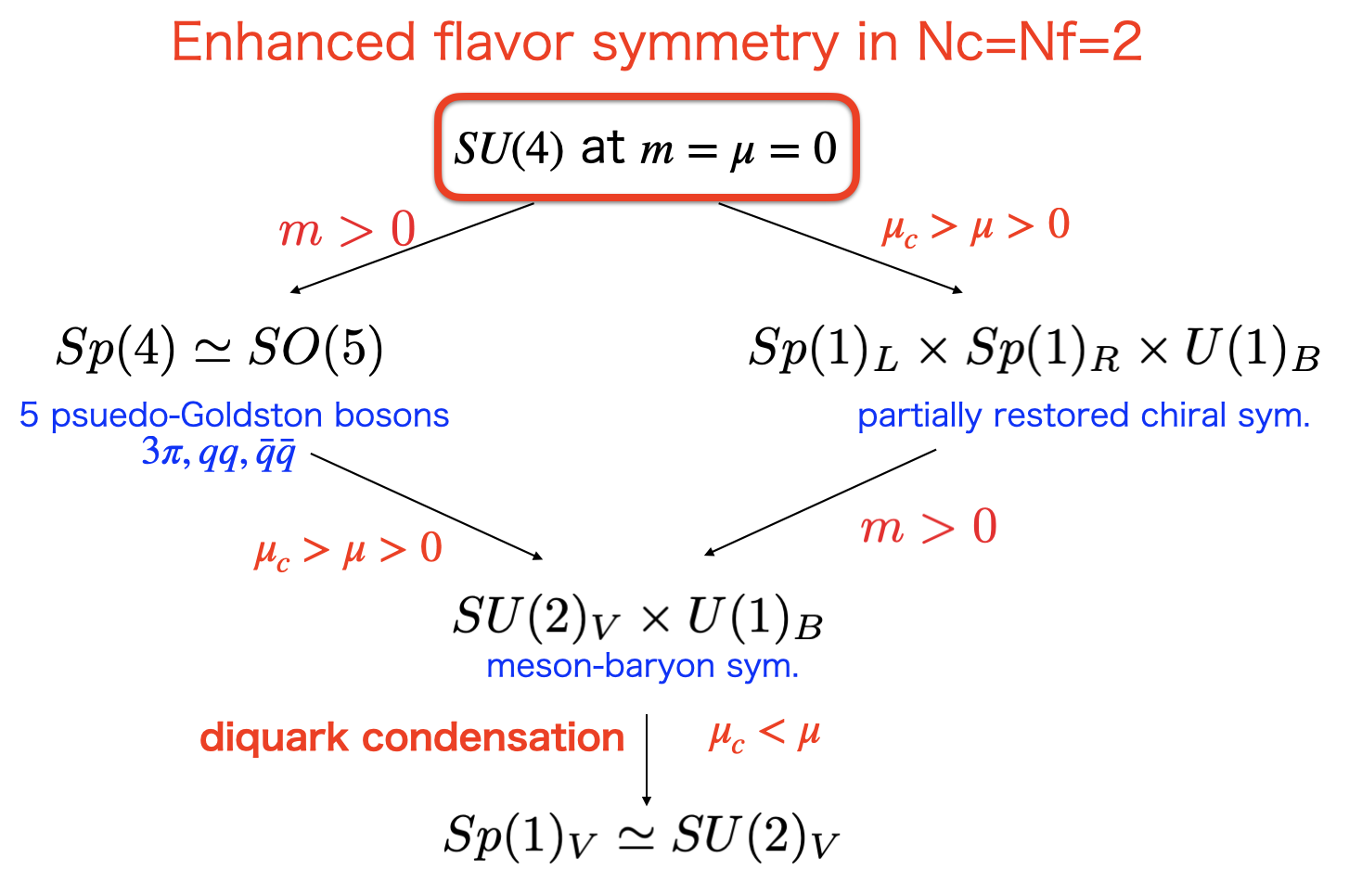}
\caption{ The flavor symmetry and its breaking structure of two-flavor QC$_2$D. } \label{fig:flavor-sym}
\end{figure}
%%%%%%%%%%%%%%%%%%%%%%%%%%%%%%%%%%%%%%%%%%%%%%%%%%%%%%%%%%
Figure~\ref{fig:flavor-sym} summarizes the flavor symmetry of two-flavor QC$_2$D.
For massless and $\mu=0$, there is a SU($2N_f$) extended flavor symmetry called the Pauli–G\"{u}rsey symmetry, which comes from the pseudo-reality of the fundamental representation in the (color) SU(2) group.
If small non-zero masses and chemical potential are put in, the $SU(2)_V \times U(1)_B$ remains as a flavor symmetry. Thus, there exists a symmetry between mesons and baryons, and the masses of the PS (or $\pi$) meson and the baryons (diquark ($qq$) and anti-diquark ($\bar{q}\bar{q}$)) are degenerate. 
Surpassing the threshold of critical value, where $\mu$ exceeds the half of the lightest baryon mass ($\mu_c =m_{qq}/2=m_{\rm PS}/2$), triggers diquark condensation and spontaneous breaking of the $U(1)_B$ baryon symmetry, resulting in a superfluid phase~\cite{Kogut:2000ek}~\footnote{In three-color QCD with an isospin chemical potential, a similar symmetry breaking for a triplet structure of $SU(2)_V$ occurs when charged pions begin to condense, breaking the third component of isospin and likewise producing superfluidity~\cite{Brandt:2017oyy}.}. 
To see such a spontaneous symmetry breaking in finite-volume simulations, a source term for explicit symmetry-breaking must be introduced into the action to probe the order parameters. Fortunately, in both dense QC$_2$D and three-color QCD with isospin chemical potential, these source terms are written by a quark-bilinear operator, making them somewhat straightforward to implement in lattice simulations.

The paper is organized as follows.
In Sec.~\ref{sec:lattice-formula}, we present the lattice formulation of QC$_2$D, including the explicit form of the lattice action and how Monte Carlo simulations are performed when the diquark source term is introduced into the action. We also summarize the simulation parameters used in our works~\cite{Iida:2019rah, Iida:2020emi,Iida:2022hyy, Ishiguro:2021yxr, Iida:2024irv}.
In Sec.~\ref{sec:phase-diagram}, we discuss the phase diagram. We provide an overview summarizing the results from various lattice groups, and then examine each physical observable, such as the diquark condensate, chiral condensate, quark number density, and confinement, presenting numerical results and corresponding figures.
Section~\ref{sec:EoS} is devoted to the EoS and the speed of sound. After providing an overview, we summarize two papers from our group~\cite{Iida:2022hyy, Iida:2024irv} and discuss the current status of related studies by other collaborations. We conclude the section with recent studies of the speed of sound relevant to neutron star physics.
In Sec.~\ref{sec:summary}, we give a summary and outlook.

%%%%%%%%%%%%%%%%%%%%%%%%%%%%%%%%%%%%%%%%%%
%%%%%%%%%%%%%%%%%%%%%%%%%%%%%%%%%%%%%%%%%%
\section{Lattice formula of QC$_2$D}\label{sec:lattice-formula}
%%%%%%%%%%%%%%%%%%%%%%%%%%%%%%%%%%%%%%%%%%
%%%%%%%%%%%%%%%%%%%%%%%%%%%%%%%%%%%%%%%%%%
Here, we summarize the lattice setup for two-flavor QC$_2$D employed in our series of studies~\cite{Iida:2019rah, Iida:2020emi,Iida:2022hyy, Ishiguro:2021yxr, Iida:2024irv}, where we use the Iwasaki gauge action with the Wilson fermion as our lattice formula. 
In the other two-flavor QC$_2$D lattice studies, various lattice formula have been adopted; using the Wilson–Plaquette gauge action with Wilson fermions~\cite{Skullerud:2003yc, Allton:2003vx, Hands:2006ec, Hands:2006ve,   Hands:2007uc, Hands:2011ye, Hands:2011hd,  Cotter:2012mb, Hands:2012fs, Boz:2015ppa, Boz:2019enj, Hands:2024nkx}, the tree-level improved Symanzik gauge action with rooted staggered fermions~\cite{Braguta:2016cpw, Astrakhantsev:2018uzd,Astrakhantsev:2020tdl, Begun:2022bxj}, and a combination either the Wilson or an improved gauge action with rooted staggered fermions~\cite{Buividovich:2020dks, Buividovich:2020gnl, Buividovich:2021fsa}. 
Although none of these studies has taken the continuum limit in a strict sense, it is nonetheless noteworthy that their results are mostly consistent.

%%%%%%%%%%%%%%%%%%%%%%%%%%%%%%%%%%%%%%%%%%%%%
\subsection{Lattice action}
As a lattice gauge action relevant to $N_c$-QCD, where $N_c$ denotes the number of colors, an improved gauge action composed of the plaquette term with $W^{1\times 1}_{\mu\nu}$ 
and the rectangular term with $W^{1\times 2}_{\mu\nu}$ is given by 
\beq
S_g = \beta \sum_x \left(
 c_0 \sum^{4}_{\substack{\mu<\nu \\ \mu,\nu=1}} W^{1\times 1}_{\mu\nu}(x) +
 c_1 \sum^{4}_{\substack{\mu\neq\nu \\ \mu,\nu=1}} W^{1\times 2}_{\mu\nu}(x) \right) 
\eeq
with $\beta=2N_c/g_0^2$, where $g_0$ denotes the bare gauge coupling constant.  
If we choose the coefficients as $c_0=1-8c_1$ with $c_1=-0.331$, then it is called the Iwasaki gauge action~\cite{Iwasaki:1983iya}.

As a lattice fermion action applicable to dense $N_c$-QCD, the two-flavor Wilson fermion action can be written as 
\beq
S_F= \bar{\psi}_1 \Delta(\mu)\psi_1 + \bar{\psi}_2 \Delta(\mu) \psi_2,\label{eq:fermion-action-wo-j}
\eeq
where the index $i=1,2$ of $\psi_i$ denotes the flavor.
Here, $\Delta(\mu)$ expresses the Wilson-Dirac operator including the number operator, 
\begin{align}
    \Delta(\mu)_{x,y} = \delta_{x,y} 
&- \kappa \sum_{i=1}^3  \left[ ( \mathbb{I}_4 - \gamma_i)  U_{x,i}\delta_{x+\hat{i},y} + (\mathbb{I}_4+\gamma_i)  U^\dagger_{y,i}\delta_{x-\hat{i},y}  \right] \nonumber\\
&- \kappa   \left[ e^{+\mu}( \mathbb{I}_4 - \gamma_4)  U_{x,4}\delta_{x+\hat{4},y} + e^{-\mu}(\mathbb{I}_4+\gamma_4)  U^\dagger_{y,4}\delta_{x-\hat{4},y}  \right],\label{eq:Dirac-op}
\end{align}
where the chemical potential is incorporated via an exponential factor, rather than a linear coefficient of $\gamma_4$, to avoid a quadratic divergence in the continuum limit~\cite{Hasenfratz:1983ba}, and $\kappa$ denotes the hopping parameter.
This $\Delta(\mu)$ breaks the $\gamma_5$-hermiticity, but still satisfies $\Delta(\mu)^\dag = \gamma_5 \Delta(-\mu) \gamma_5$. In the lattice Monte Carlo simulation, the fermion action is expressed as $\det (\Delta (\mu))$ by integrating out the fermion fields beforehand.
At $\mu \ne 0$, 
\beq
(\det \Delta (\mu))^* = \det \Delta (\mu) ^\dag = \det \Delta (-\mu) \label{eq:non-hermicity}
\eeq
implies that the fermion action takes a complex value in general.
On the other hand, we can easily see the two-flavor system of $N_c$-QCD with a degenerated mass under the isospin chemical potential $\mu_d= - \mu_u$ satisfies the real-positivity of the fermion action, namely $[\det \Delta (\mu_u) \det \Delta (\mu_d) ]^\dag = [\det \Delta (\mu_u) \det \Delta (\mu_d)]$.

In the case of $N_c=2$, namely QC$_2$D, the link variable, $U_\mu= e^{i A_\mu^a \tau^a}$, satisfies
\beq
U_{\mu}^* = \tau_2 U_\mu \tau_2,
\eeq
where $\tau_2$ denotes the Pauli matrix and acts on the color indices.
Then, for given flavor-blind $\mu$, the Dirac operator has the following conjugate property:
\beq
\Delta (\mu)^* = \tau_2 ( C \gamma_5) \Delta (\mu) (C \gamma_5)^{-1} \tau_2,
\eeq
where $C \gamma_5$ acts on the spinor indices, and $C$ denotes the charge conjugation operator; here we take $C=i \gamma_0 \gamma_2$.
Consequently,  
one obtains $[\det \Delta (\mu)]^* = \det \Delta (\mu)$.
Thus, the fermion action in SU($2$) gauge theory takes a real value.
Essentially, this comes from the pseudo-reality of the fundamental representation, namely ``quarks'', of the SU($2$) group.
Note that here the positivity of the fermion action is not guaranteed. If we consider an odd-number fermion system, the action can take a real but negative value, and hence the sign problem appears even in QC$_2$D~\cite{Fukushima:2008su}.

To perform the low-temperature and high-density QC$_2$D simulations, such a positive-reality of the fermion action is necessary but insufficient.
In general QCD(-like) theory, it is known that a numerical instability related to the (early-)onset problem appears around $\mu = m_{\rm PS}/2$~\cite{Muroya:2000qp,Cohen:2003kd, Muroya:2003qs}. In the case of QC$_2$D, in such a $\mu$ regime, there is a phase transition from the hadronic (normal) to the superfluid phase.
Then, the diquark mass approaches zero around the phase transition that the simulation could not proceed beyond $\mu \approx m_{\rm PS}/2$ even if we utilized a very tiny molecular-dynamics step in the Hybrid Monte Carlo algorithm, for instance, the step-size $\sim 1/1000$ could not work for $16^4$ lattice in our simulations~\cite{Iida:2019rah}.

To avoid this problem, we introduce the diquark source term to the fermion action on the lattice,
\beq
S_F= \bar{\psi}_1 \Delta(\mu)\psi_1 + \bar{\psi}_2 \Delta(\mu) \psi_2 - J \bar{\psi}_1 (C \gamma_5) \tau_2 \bar{\psi}_2^{T} + \bar{J} \psi_2^T (C \gamma_5) \tau_2 \psi_1,\label{eq:action}
\eeq
with $J=j \kappa$ and $\bar{J}= \bar{j} \kappa$ corresponding to the anti-diquark and diquark source parameters, respectively. Here, the factor $\kappa$ comes from the rescaling of the Wilson fermion on the lattice. 
For simplicity, we put $J=\bar{J}$ and assume that it takes a real value. 
The term serves as an external source that explicitly breaks the U(1) symmetry relating the diquark and anti-diquark. Spontaneous symmetry breaking can then be diagnosed by extrapolating the source strength $j \to 0$ and examining whether the expectation value of the associated operator remains nonzero.

The question is how to incorporate this action into the HMC framework and run the simulation.
Indeed, the action~\eqref{eq:action} still has a bilinear form of fermions.
It can be rewritten by using an extended fermion matrix ($\mathcal M$) as
\beq
S_F&=& (\bar{\psi}_1 ~~ \bar{\varphi}) \left( 
\begin{array}{cc}
\Delta(\mu) & J \gamma_5 \\
-J \gamma_5 & \Delta(-\mu) 
\end{array}
\right)
\left( 
\begin{array}{c}
\psi_1  \\
\varphi  
\end{array}
\right)
 \equiv  \bar{\Psi} {\mathcal M} \Psi,  \label{eq:def-M}
\eeq
where
$\bar{\varphi}=-\psi_2^T C \tau_2, ~~~ \varphi=C^{-1} \tau_2 \bar{\psi}_2^T.$
Thanks to the pseudo-reality of the fundamental fermions in the SU($2$) gauge theory, $\psi_1$ and the charge conjugation of $\bar{\psi}_2^T$ can be put in the same multiplet. 
The square of the extended matrix can be diagonal if $J(=\bar{J})$ is real.   
We thus obtain
\beq
\det[{\mathcal M}^\dag {\mathcal M}] = 
\det[ \Delta^\dag(\mu)\Delta(\mu) + J^2 ]~\det[  \Delta^\dag (-\mu) \Delta(-\mu) + J^2  ]. \label{eq:MdagM}
\eeq
From the point of view of actual numerical calculations, the $J$ insertion speeds up calculations since it lifts the eigenvalues of the matrix up~\cite{Kogut:2001na, Kogut:2002cm,Skullerud:2003yc}.

In fact, $\det[{\mathcal M}^\dag {\mathcal M}]$ corresponds to the fermion 
action for the four-flavor theory, since a single $\mathcal{M}$ in 
Eq.~\eqref{eq:def-M} represents the fermion kernel of the two-flavor 
theory.  To reduce the number of flavors, we take the square root of the extended matrix in the action and utilize the Rational Hybrid Monte Carlo (RHMC) algorithm in our numerical simulations.

A similar approach can be applied to three-color QCD with an isospin chemical potential by including a pionic source term in the lattice action. Because this pionic source is also a fermion bilinear operator, $\pi^{\pm}=\bar{\psi}_1 \gamma_5 \psi_2 - \bar{\psi}_2 \gamma_5 \psi_1$, it can be implemented within a similar framework of the RHMC algorithm. For details of the implementation, see Ref.~\cite{Brandt:2017oyy}.

%%%%%%%%%%%%%%%%%%%%%%%%%%%%%%%%%%%%%%%%%%%%%
\subsection{Simulation setup}
In this subsection, we summarize the lattice setup for our data, which will be mainly shown in this manuscript.
We have carried out HMC simulations of QC$_2$D using the Iwasaki gauge action and unimproved Wilson fermions extended by a quark chemical potential term at $j=0$ in low-$\mu$ regimes. At high density, a diquark source term \(j\), as shown in the previous subsection, must also be included in the action. In the regime, we have performed the RHMC simulations to generate configurations.

We fix the ``physical'' scale so that the scale correspondence is easy to understand, though QC$_2$D is not a real theory.
According to Ref.~\cite{Iida:2020emi}, in our lattice action, the chiral susceptibility at $\mu=0$ 
 is peaked at $(\beta, N_\tau$) $=$ ($0.950,10$) with the simulations performed at fixed $m_{\rm PS}/m_{V}$~\footnote{The quantity corresponds to $m_{\pi}/m_\rho$ in three-color QCD.}, where $m_{\rm PS}$  and $m_V$ are the pseudo-scalar and vector meson masses in vacuum ($\mu=0$), respectively. 
Using the scale setting with $w_0$~\cite{BMW:2012hcm} scale in the gradient flow method~\cite{Luscher:2010iy}, we fix the physical lattice spacing as $a=0.17$ fm if we set $T_c = 200$ MeV to fix a physical unit~\cite{Iida:2020emi}.

All runs we shown in this manuscript employ \((\beta,\kappa)=(0.800,0.159)\), with three lattice sizes; \(32^4\) (\(T=40\) MeV), \(16^4\) (\(T=80\) MeV) and \(32^3\times8\) (\(T=160\) MeV). 
In the vacuum study at the same coupling and hopping parameters~\cite{Iida:2020emi, Murakami:2022lmq,Murakami:2023ejc}, we found
\[
\frac{m_{\rm PS}}{m_V}=0.813(1), 
\quad a m_{\rm PS}=0.620(1)\ (\,m_{\rm PS}\approx738\ \mathrm{MeV}\,).
\]
For our main results in $32^4$ lattices (T = 40MeV), we scan the dimensionless chemical potential up to \(a\mu=0.75\) in steps of \(\Delta a\mu=0.05\), where we keep \(a\mu\le0.75\) to avoid lattice artifact phases. 
Furthermore, we also perform the simulations for $a\mu =0.27$, which is near the onset scale, $\mu/m_{\rm PS}=0.5$.
For \(a\mu<0.27\) we simulate at \(j=0\), while for \(a\mu\ge0.27\) we use \(a j=0.010,\,0.015,\,0.020\) and extrapolate each observable to \(j\to0\) at fixed $a\mu$. At each parameter set, we generate 50 -- 100 configurations and estimate statistical errors by the jackknife method.
For detailed lattice parameters on $16^4$ and $32^3\times 8$ lattices, see Ref.~\cite{Iida:2019rah}.

%%%%%%%%%%%%%%%%%%%%%%%%%%%%%%%%%%%%%%%%%%
\section{Phase diagram}\label{sec:phase-diagram}
%%%%%%%%%%%%%%%%%%%%%%%%%%%%%%%%%%%%%%%%%%%%%
%%%%%%%%%%%%%%%%%%%%%%%%%%%%%%%%%%%%%%%%%%%%%
\subsection{Overview}
\begin{figure}[h]
\centering
\includegraphics[width=1.0 \textwidth]{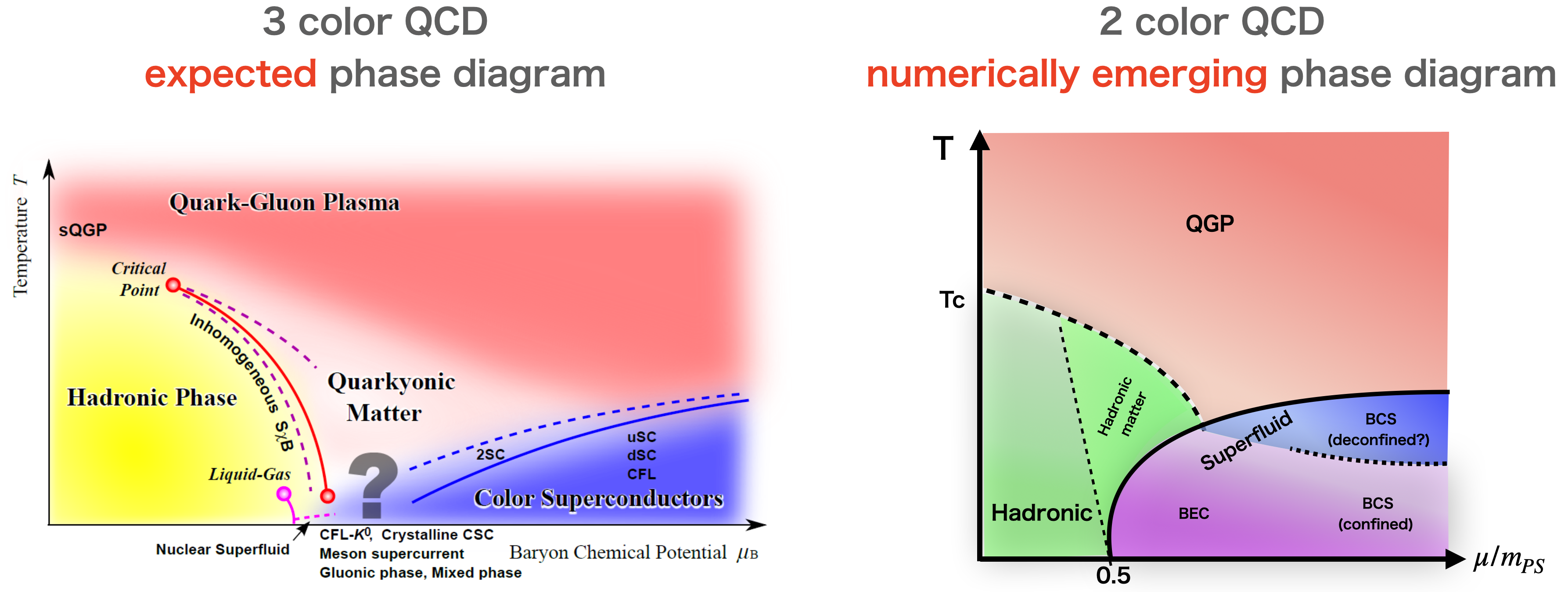}
\caption{Comparison between three-color and two-color QCD phase diagrams. Left (Right) figures are originally shown in Ref.~\cite{Fukushima:2010bq} (\cite{Iida:2024irv}). } \label{fig:comp-phase-diagram}
\end{figure}

The phase diagram of three-color QCD is generally expected to consist of three distinct regions: the hadronic phase at low temperature and low baryon density, the quark-gluon plasma (QGP) phase at high temperature, and the color-superconducting phase at low temperature and high density. 
The extreme regions of the phase diagram are well understood by analytical studies, and the intermediate-temperature region at zero chemical potential (i.e., along the vertical axis) has been extensively studied through lattice simulations, revealing the nature of the thermal phase transition (which is, in fact, a crossover rather than a true phase transition). In contrast, the low-temperature and intermediate-density region remains poorly understood, as the sign problem hampers first-principles calculations in this regime.
Based on this situation, for example, a QCD phase diagram is expected as the one in the left panel of Figure~\ref{fig:comp-phase-diagram}.

%%%%%%%%%%%%%%%%%%%%%%%%%%%%%%%%%%%%%%%%%%%%%%%%%%%%%%%%%%
%%%%%%%%%%%%%%%%%%%%%%%%%%%%%%%%%%%%%%%%%%%%%%%%%%%%%%%%%%

In contrast, QC$_2$D is free of the sign problem, and we can handle the onset problem even at a high chemical potential region, as explained in the previous section. Thanks to recent lattice studies, a qualitatively consistent picture has emerged, as shown in the right panel of Figure~\ref{fig:comp-phase-diagram}.
Indeed, several independent lattice studies have revealed the following features of the QC$_2$D phase diagram.
First, the superfluidity phase transition occurs around $\mu_c \approx m_{\rm PS}/2$, and the value of $\mu_c$ shows only a weak temperature dependence.
Second, the superfluid phase persists up to relatively high temperatures, as large as $T \approx 100$~MeV.
Third, the superfluid phase itself encompasses two distinct regimes: at lower densities, the system exhibits Bose–Einstein condensation (BEC)–like behavior with tightly bound diquark pairs, in accordance with predictions from chiral perturbation theory (ChPT); while at higher densities, it goes to a Bardeen–Cooper–Schrieffer (BCS)–like regime, where the diquark pairing is weakly bound and consistent with perturbative analyses of QCD.
This crossover from BEC to BCS is a hallmark of strongly interacting fermionic systems and provides further insight into the nature of dense matter in QCD-like theories. Furthermore, the low-temperature, high-density regime (even at $\mu \approx 1$GeV) remains in a confined phase in QC$_2$D, in contrast to the expected deconfined nature of dense matter in three-color QCD, where quark degrees of freedom are anticipated to dominate on a macroscopic scale.
This difference between two-color and three-color QCD is expected to originate from the nature of the diquark condensate that acquires a vacuum expectation value in the superfluid phase, being a color singlet in QC$_2$D, whereas a colored quantity in three-color QCD.

Although these findings have yet to be independently confirmed, several recent studies have also suggested the existence of a so-called hadronic-matter phase at finite temperature, located just below the superfluid phase-transition point~\cite{Iida:2019rah, Iida:2024irv}. This phase is characterized by a non-zero quark number density and vanishing diquark condensate. It has been observed that the extent of this phase shrinks as the temperature decreases, indicating that it is driven by thermal excitations.
In addition, it has been observed that confined phases at high density can support nontrivial topological configurations, with the topological susceptibility showing little dependence on chemical potential~\cite{Iida:2019rah, Iida:2024irv}. This suggests that, although the low-temperature and high-density regimes appear to be consistent with perturbative QCD in terms of quark behavior, the gluonic sector may remain nonperturbative, unlike in the high-temperature regime.
Moreover, several recent works~\cite{Boz:2019enj, Begun:2022bxj} suggest that confinement may persist at all values of $\mu$.

In the following sections, we provide detailed discussions of these developments in the study of dense QC$_2$D, highlighting both established results and ongoing open questions.

%%%%%%%%%%%%%%%%%%%%%%%%%%%%%%%%%%%%%%%%%%%%%
\subsection{Observables and definition of phases}
%%%%%%%%%%%%%%%%%%%%%%%%%%%%%%%%%%%%%%%%%%%%%
Now, let us make clear the notation of the phase.
We use the name of each phase as shown in Table~\ref{table:phase}.
In this manuscript, we utilize three quantities, the magnitude of the Polyakov loop ($\langle |L| \rangle$), the diquark condensate ($\langle qq \rangle$), and the quark number density ($\langle n_q \rangle$) to distinguish among different phases.
%%%%%%%%%%%%%%
\begin{table}[h]
\begin{center}
\caption{ Definition of the phases. To distinguish between the BEC and BCS phases, we use the value of $\langle n_q \rangle$. Meanwhile, it is expected that $\langle qq \rangle$ scales as  $\propto\mu^2$ by the weak coupling analysis~\cite{Schafer:1999fe, Hanada:2011ju, Kanazawa:2013crb}. } \label{table:phase}
\begin{tabular}{|c||c|c|c|c|c|}
\hline
 \multicolumn{1}{|c||}{}  & \multicolumn{2}{c|}{Hadronic } &     \multicolumn{2}{c|}{Superfluid } &\multicolumn{1}{c|}{QGP} \\  
\cline{3-3} \cline{4-5}  & & Hadronic matter ($T>0$) &BEC & BCS  &  \\  
 \hline \hline
$\langle |L| \rangle$ & zero  & zero  &   &   & non-zero \\
$\langle qq \rangle$ & zero  &  zero  & non-zero & ($\propto \mu^2$)  & zero \\ 
$ \langle n_q \rangle $ &   zero &  non-zero  & non-zero & $\langle n_q \rangle /n_q^{\rm tree} \approx 1$ & non-zero \\ 
 \hline
\end{tabular}

\end{center}
\end{table}
%%%%%%%%%%%%%
Here, we provide the definition of these observables and the chiral condensate, $\langle \bar{q}q \rangle$, and the topological susceptibility, which are also interesting quantities to see the vacuum properties for each phase.

Firstly, the Polyakov loop is given by
\beq
L= \frac{1}{N_s^3} \sum_{\vec{x}} \prod_\tau U_{4} (\vec{x}, \tau),
\eeq
which plays the role of an approximate order parameter for confinement.
As we will explain in Sec.~\ref{sec:confinement}, it has been found that this quantity is often affected by severe finite-volume effects. Therefore, it is also important to examine the quark-antiquark potential to see if confinement occurs.
Secondly, the diquark condensate,
\beq
\langle qq \rangle \equiv \frac{\kappa}{2} \langle \bar{\psi}_1 (C \gamma_5) \tau_2 \bar{\psi}_2^T - \psi_2^T (C \gamma_5) \tau_2 \psi_1 \rangle,\label{eq:def-qq}
\eeq 
is the order parameter for superfluidity, which is an associated operator of the external source of the U(1)$_B$ breaking term ($j$) in Eq.~\eqref{eq:action}. We refer to the regime with $\langle qq \rangle \ne 0$ in the $j \to 0$ limit as the superfluid phase.
The third quantity, namely, the quark number density,
\beq
a^3 n_q= \sum_{i} \kappa \langle \bar{\psi}_i (x) (\gamma_4 -\mathbb{I}_4) e^\mu U_{4} (x) \psi_i (x+\hat{4})  + \bar{\psi}_i (x) (\gamma_4 + \mathbb{I}_4) e^{-\mu}U_4^\dag (x-\hat{4} )\psi_i (x-\hat{4}) \rangle,\nonumber\\
\eeq
is utilized to identify a BEC-BCS crossover behavior in the superfluid phase.
This quantity is the time-like component of a conserved current and does not require a renormalization.
It goes to $2N_fN_c$ in the high-density limit on the lattice.

It might be worthwhile to see the chiral condensate, 
\beq
\langle \bar{q} q \rangle \equiv \langle \bar{\psi}_i \psi_i \rangle,
\eeq
which is well-known as an order parameter of chiral symmetry breaking. Here, we define the chiral condensate for one flavor and do not take the sum over the flavor index $i$.  In our numerical calculations, we utilize the massive Wilson fermions, which explicitly break the chirality.  Although we will discuss numerical results for the chiral condensate in each phase, we will not use the quantity for the definition 
of the phase.

Furthermore, studying possible classical configurations by observing the topological-charge distribution is also interesting. 
The topological charge can be measured by incorporating the gluonic definition,
\beq
Q (t) = \frac{1}{32 \pi^2} \sum_{x} \mbox{Tr} \epsilon_{\mu \nu \rho \sigma} 
G^a_{\mu \nu} (x,t) G^a_{\rho \sigma} (x,t),
\eeq
into the gradient flow method~\cite{Luscher:2010iy}.  Here, the field strength $G_{\mu \nu}(x,t)$ is constructed by the smeared link variables at finite flow-time $t$. The field strength is calculated by using the clover-leaf operator on the lattice.

The value of $Q(t)$ roughly plateaus for long $t$, but small fluctuations remain. Therefore, we introduce a reference scale $t_0$ and identify the value of $Q(t=t_0)$ as a convergent value of $Q$ for each configuration~\cite{Bruno:2014ova}.
The reference scale $t_0$ is originally introduced in Ref.~\cite{Luscher:2010iy}, it is given by
\beq
t^2 \langle E(t) \rangle |_{t=t_0} =0.3,\label{eq:t0-scale}
\eeq
where $E(t)$ denotes the energy density
\beq
E(t)= -\frac{1}{2V} \sum_x \mathrm{tr} \{ G_{\mu \nu} (x,t) G_{\mu \nu}(x,t) \}.
\eeq
Finally, the distribution of the topological charge, namely the topological susceptibility, is evaluated by
\beq
\chi_Q = \langle Q^2 \rangle - \langle Q \rangle^2.\label{eq:def-chi_Q}
\eeq

%%%%%%%%%%%%%%%%%%%%%%%%%%%%%%%%%%%%%%%%%%%%%
\subsection{Hadronic-superfluid phase transition}\label{sec:hadron-SF}
%%%%%%%%%%%%%%%%%%%%%%%%%%%%%%%%%%%%%%%%%%%%%
At low temperature and high density, it is expected that the system undergoes a transition to a superfluid phase (or color superconducting phase). The order parameter for this transition is the diquark condensate, $\langle qq \rangle$ given in Eq.~\eqref{eq:def-qq}. In three-color dense QCD, this condensate carries color-charge and hence the phase is referred to as the color-superconducting phase (see the left panel of Figure~\ref{fig:comp-phase-diagram}). In contrast, in QC$_2$D, the diquark condensate is color-singlet. 
Similarly, in three-color QCD with a finite isospin chemical potential, pion condensation occurs in the corresponding region. Since the pion is also a color singlet, this phase exhibits superfluidity as well.

In the case of QC$_2$D at zero temperature, the mean-field ChPT analysis~\cite{Kogut:2000ek} and other chiral effective models~\cite{Ratti:2004ra} predict that the diquark condensation occurs when the chemical potential $\mu$ reaches approximately $m_{\rm PS}/2$, which also corresponds to the so-called onset scale. Near the critical point, the order parameter is expected to scale as $\langle qq \rangle \sim (\mu - \mu_c)^{1/2}$~\cite{Kogut:2000ek}.
Many independent lattice studies have confirmed both the onset value of $\mu$ and the critical exponent of $1/2$ predicted by the scaling law even at finite but sufficiently low temperatures~\cite{Hands:2006ve, Braguta:2016cpw, Boz:2019enj, Iida:2019rah}. 

Using this fact, in our paper~\cite{Iida:2024irv}, where we study $T=40$MeV, to find the critical $\mu_c$ where the hadronic-superfluid phase transition occurs, we attempt to fit a few data for $\langle qq \rangle$ right after the diquark condensate becomes non-zero by using the scaling law,
\beq
\langle qq \rangle =A (\mu -B)^{1/2},\label{eq:qq-ChPT-scaling}
\eeq
with the fitting parameters, $A$ and $B$.
The fit yields $B$ which corresponds to $\mu_c$.

%%%%%%%%%%%%%%%%%%%%%%%%%%%%%%%%%%%%%%%%%%%%%%%%%%%%%%%%%%
\begin{figure}[h]
\centering
\includegraphics[width=0.4 \textwidth]{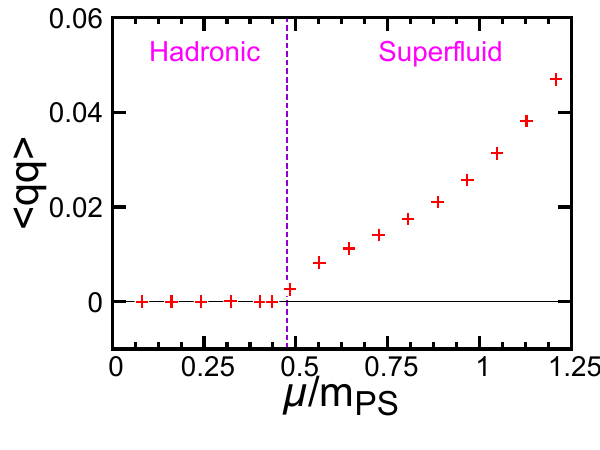}
\includegraphics[width=0.4 \textwidth]{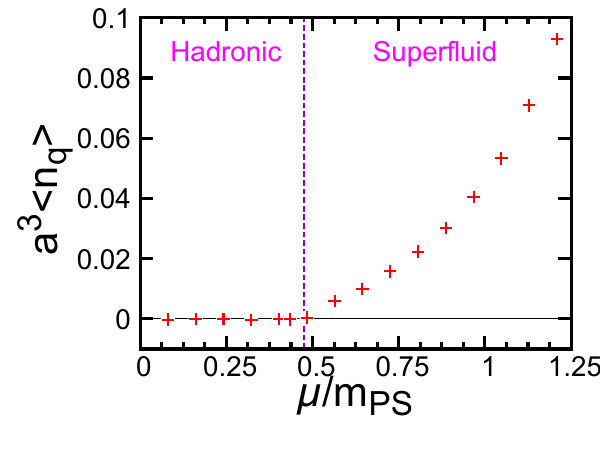}
\caption{(Left) diquark condensate as a function of $\mu$. (Right) quark number density in lattice unit as a function of $\mu$. The data are originally obtained in Ref.~\cite{Iida:2024irv}.} \label{fig:diquark-cond-and-nq}
\end{figure}
%%%%%%%%%%%%%%%%%%%%%%%%%%%%%%%%%%%%%%%%%%%%%%%%%%%%%%%%%%
The left panel of Figure~\ref{fig:diquark-cond-and-nq}  depicts the lattice results obtained in Ref.~\cite{Iida:2024irv}. It clearly shows that the diquark condensate $\langle qq \rangle$ becomes nonzero at $\mu \sim m_{\rm PS}/2$ and then increases with increasing $\mu$. The location of the hadronic-superfluid phase transition, $\mu_c$, is determined to be around $\mu/m_{\rm PS} = 0.48$, based on a fit to the data points near the onset region.

It is also known that $\langle qq \rangle$ eventually decreases at very high density. This behavior was demonstrated in Ref.~\cite{Kogut:2001na} for the case of $N_f = 4$, and it is attributed to lattice artifacts that become significant when $a\mu \sim 1$, due to the influence of the lattice cutoff.
In general, there are two known sources that the upper-limit on $\mu$: one is the lattice artifact arising from large $a\mu$, and the other is the saturation effect, where all lattice sites are filled with particles. In practice, the former typically imposes a lower upper-limit on $\mu$.
In our lattice setup, the upper-limit is estimated as $a\mu \approx 0.80$ ($\mu /m_{\rm PS} \approx 1.25$) from Figure~6 in Ref.~\cite{Iida:2019rah}.

As summarized in Table~\ref{table:phase}, another characteristic feature of the superfluid phase is the spontaneous breaking of the U(1) baryon symmetry, which results in a nonzero vacuum expectation value of the quark number density $\langle n_q \rangle$. This quantity represents the difference between the number of quarks and antiquarks.

Although the location of the hadronic–superfluid phase transition was determined by examining the order parameter $\langle qq \rangle$, we also observe that $\langle n_q \rangle$ becomes non-zero at almost the same value of the chemical potential in the right panel of Figure~\ref{fig:diquark-cond-and-nq}.
A more detailed discussion related to the hadronic-matter phase emerged at higher temperature, $T=80$MeV, where $\langle qq \rangle =0$ but $\langle n_q \rangle >0$, is given in  Section 3.2 in Ref.~\cite{Iida:2024irv}.

%%%%%%%%%%%%%%%%%%%%%%%%%%%%%%%%%%%%%%%%%%%%%
\subsection{BEC-BCS crossover}\label{sec:BEC-BCS}
\begin{figure}[htbp]
\centering
\includegraphics[width=.9\textwidth]{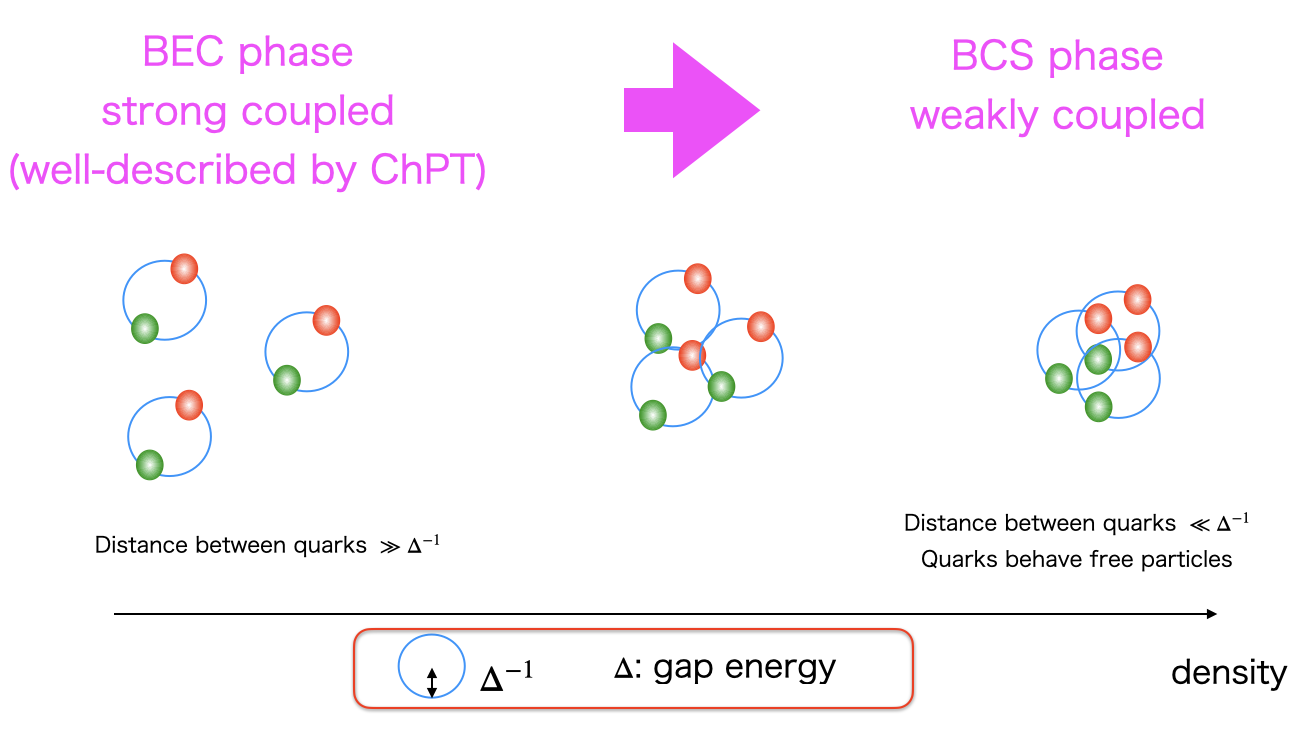}
\caption{A schematic picture of the BEC–BCS crossover in dense QC$_2$D. In this figure, the gap energy is illustrated as being independent of $\mu$; however, note that the actual $\mu$-dependence of the gap remains unclear.} \label{fig:image-BEC-BCS}
\end{figure}

%%%%%%%%%%%%%%%%%%%%%%%%%%%%%%%%%%%%%%%%%%%%%
In the superfluid phase, it is expected that a macroscopic number of diquarks form a condensate in such a way that the BEC-BCS crossover occurs as the density increases.
In the BEC regime, the attraction between quarks within each pair is sufficiently strong to form tightly bound diquarks. In contrast, in the BCS regime, quark dynamics can be treated as a weakly interacting system near the Fermi surface, where the instability toward Cooper pair formation arises as a small correction to the free theory. In QCD-like theories, the asymptotic freedom ensures that the quark-quark interaction becomes weaker at shorter distances. As a result, the BCS regime naturally appears at sufficiently high densities where the typical interquark separation becomes small, and the system approaches the free limit (see Figure~\ref{fig:image-BEC-BCS}).
%%%%%%%%%%%%%%%%%%%%%%%%%%%%%%%%%%%%%%%%%%%%%%%%%%%%%%%%%%
%%%%%%%%%%%%%%%%%%%%%%%%%%%%%%%%%%%%%%%%%%%%%%%%%%%%%%%%%%

The quark number density is often examined in order to identify the BCS phase~\cite{Hands:2006ve}.
As a criterion, if the lattice result $\langle n_q \rangle$ can be approximated by its value for free theory on the lattice,
\beq
n_q^{\rm tree}(\mu) = \frac{4N_cN_f}{N_s^3 N_\tau} \sum_k \frac{i \sin \tilde{k}_0 [ \sum_i \cos k_i -\frac{1}{2\kappa} ]}{[\frac{1}{2\kappa} -\sum_\nu \cos \tilde{k}_\nu ]^2 +\sum_\nu \sin^2 \tilde{k}_\nu},\label{eq:nq-tree}
\eeq
where
\beq
\tilde{k}_0 = k_0 -i\mu = \frac{2\pi}{N_\tau} (n_0+1/2) -i\mu,~~~~~~~~\tilde{k}_i = k_i = \frac{2\pi}{N_s}n_i,~~~~i=1,2,3,
\eeq
then it refers to such a regime as the BCS phase.

%%%%%%%%%%%%%%%%%%%%%%%%%%%%%%%%%%%%%%%%%%%%%%%%%%%%%%%%%%
\begin{figure}[htbp]
\centering
\includegraphics[width=.4\textwidth]{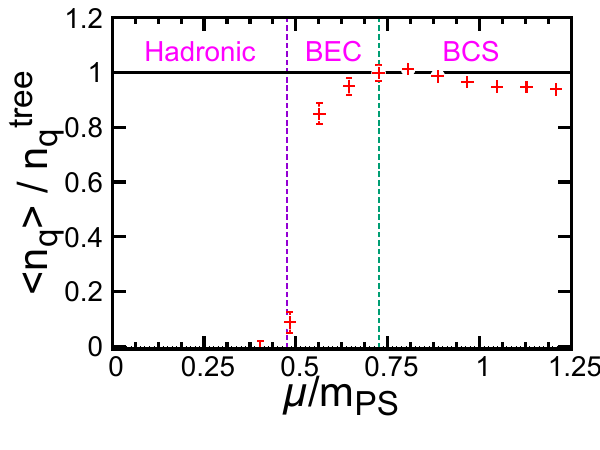}
\caption{The $\mu$ dependence of the quark number density normalized by the free theory calculation (Eq.~\eqref{eq:nq-tree}). It is originally shown in Ref.~\cite{Iida:2024irv}. \label{fig:number-density}}
\end{figure}
%%%%%%%%%%%%%%%%%%%%%%%%%%%%%%%%%%%%%%%%%%%%%%%%%%%%%%%%%%
Figure~\ref{fig:number-density} shows our results of the quark number density normalized by the tree level (free quark) calculations, Eq.~\eqref{eq:nq-tree}.
The normalized quantity reaches unity at $\mu/m_{\rm PS}=0.73$ ($a\mu=0.45$) in this plot. 
Therefore, we refer to the regime of $\mu/m_{\rm PS} \geq 0.73$ as the BCS phase, and put a guiding line at $\mu/m_{\rm PS}=0.73$ as a typical BEC-BCS crossover point in our lattice setup as shown in Refs.~\cite{Iida:2019rah, Iida:2024irv}.

Now, while the BEC–BCS crossover point is defined here from the data of the quark number density, one may ask whether the diquark condensate $\langle qq \rangle$ exhibits any qualitative change across the two regions.
In the continuum limit, Eq.~\eqref{eq:nq-tree} scales as $n_q^{\rm tree,cont}=N_c N_f \mu^3/(3\pi^2)$ as a function of $\mu$ in the high-density limit where the Fermi surface with radius $\mu$ is perfectly constructed.  
According to the weak coupling analysis~\cite{Schafer:1999fe, Hanada:2011ju, Kanazawa:2013crb}, it is expected that the diquark condensate increases as $\langle qq \rangle \propto \mu^2$ in such a high-density regime.

%%%%%%%%%%%%%%%%%%%%%%%%%%%%%%%%%%%%%%%%%%%%%%%%%%%%%%%%%%
\begin{figure}[h]
\centering
\includegraphics[width=1.0 \textwidth]{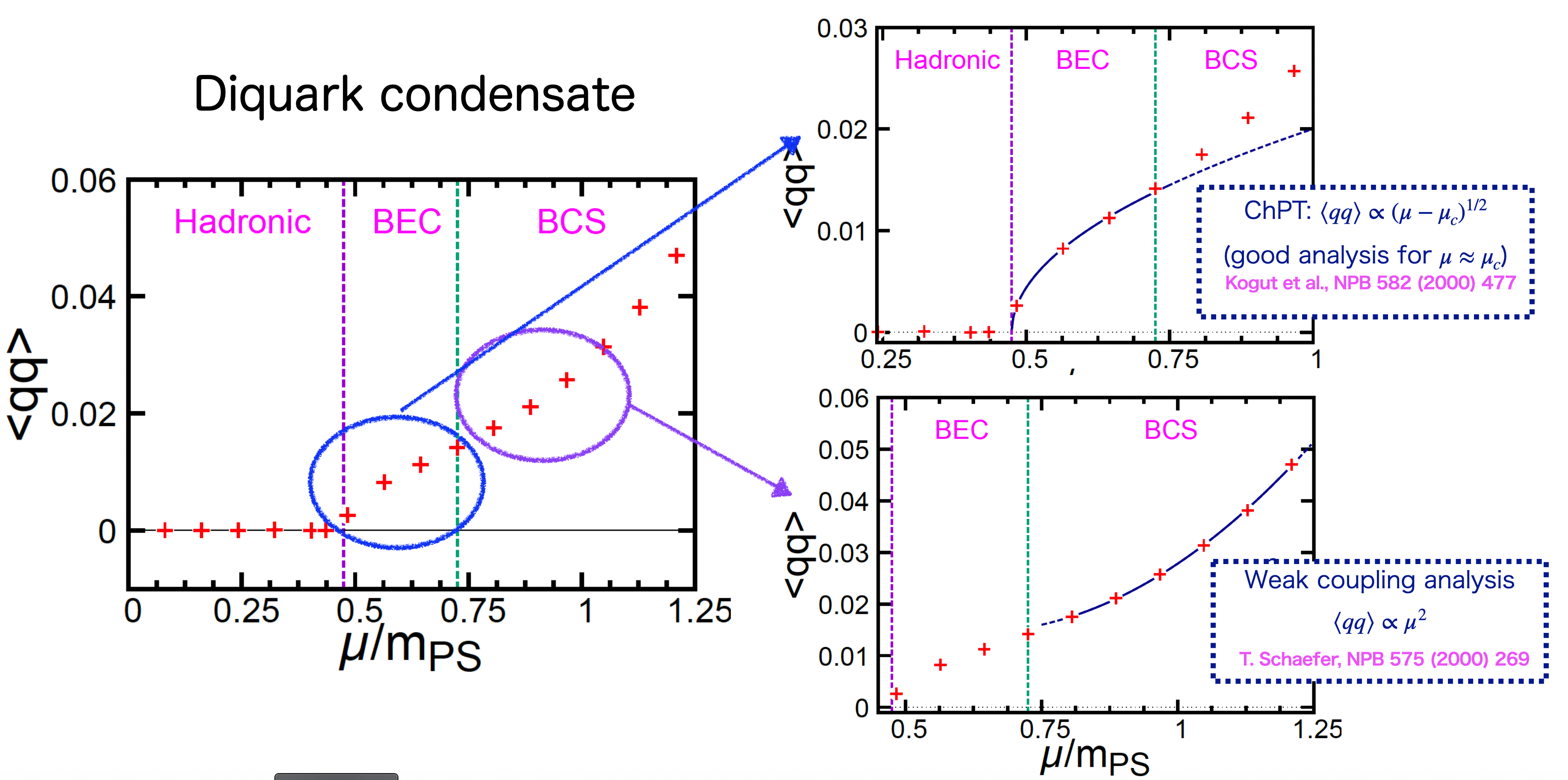}
\caption{Scaling behavior of diquark condensate in BEC and BCS phases. The two right panels are enlarged plots for the BEC (top) and BCS (bottom) regimes. The fitting functions (blue curves) are $\langle qq \rangle =A(\mu-B)^{1/2}$ in the top panel and $\langle qq \rangle = c_2 \mu^2 + c_1 \mu +c_0$ in the bottom panel, which are predicted by the ChPT and weak coupling analysis, respectively. The data are originally obtained in Ref.~\cite{Iida:2024irv}.} \label{fig:scaling-diquark-cond}
\end{figure}
%%%%%%%%%%%%%%%%%%%%%%%%%%%%%%%%%%%%%%%%%%%%%%%%%%%%%%%%%%
The two right panels of Figure~\ref{fig:scaling-diquark-cond} show the fit results of the data $\langle qq \rangle$ in the BEC and BCS phases: In the upper panel, we use the prediction of the ChPT,  $\langle qq \rangle \sim (\mu - \mu_c)^{1/2}$~\cite{Kogut:2000ek},  as a fit function in the BEC phase, while in the lower panel, we employ a quadratic function of $\mu$ in the BCS phase following the weak coupling analysis~\cite{Schafer:1999fe, Hanada:2011ju, Kanazawa:2013crb}. Although the two fitting functions exhibit clearly different curvatures with respect to $\mu$, both describe the data well in their respective regions defined by the quark number density. This supports the existence of two distinct phases.

%%%%%%%%%%%%%%%%%%%%%%%%%%%%%%%%%%%%
\subsection{Chiral condensate}\label{sec:chiral-cond}
%%%%%%%%%%%%%%%%%%%%%%%%%%%%%%%%%%%%
Now, we briefly comment on the chiral symmetry. In the case of massless fermions, the chiral symmetry is expected to be restored in the high-density superfluid phase. Thus, $\langle \bar{q}q\rangle \rightarrow 0$ is expected in the high-density regimes. In our simulations, the Wilson fermions are used, and hence the chiral symmetry is explicitly broken at the lattice level. Nevertheless, remnants of its restoration can still be observed.

%%%%%%%%%%%%%%%%%%%%%%%%%%%%%%%%%%%%%%%%%%%%%%%%%%%%%%%%%%
\begin{figure}[h]
\centering
\includegraphics[width=0.6 \textwidth]{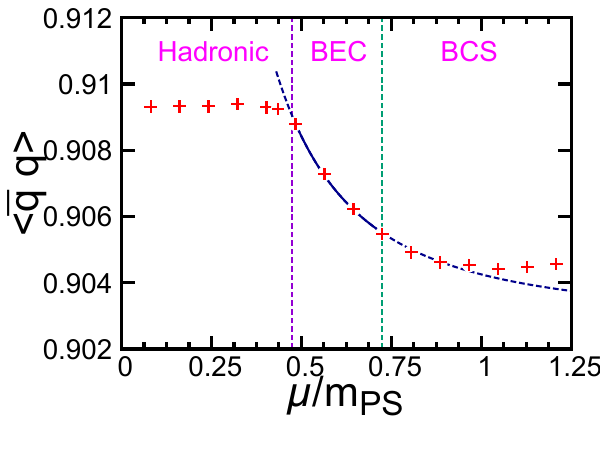}
\caption{The $\mu$ dependence of the chiral condensate in the $j=0$ limit. The blue curve presents the fitting curve using $f(\mu)=c_1(\mu_c^2/\mu^2) +c_0$. It is originally shown in Ref.~\cite{Iida:2024irv}. } \label{fig:scaling-chiral-cond}
\end{figure}
%%%%%%%%%%%%%%%%%%%%%%%%%%%%%%%%%%%%%%%%%%%%%%%%%%%%%%%%%%
Figure~\ref{fig:scaling-chiral-cond} shows the $\mu$-dependence of the chiral condensate.  In the hadronic phase, the value of the chiral condensate remains unchanged as $\mu$ is varied. However, once the system undergoes a transition into the superfluid phase, the condensate starts to decrease. This decrease in the BEC region is well fitted by the scaling law predicted by ChPT,
\begin{equation}
\langle \bar{\psi} \psi \rangle (\mu) \propto \frac{\mu_c^2}{\mu^2}.
\label{eq:chpt_scaling}
\end{equation}
In the BCS phase, the condensate becomes nearly constant, forming a plateau. Although the Wilson fermions require an additive renormalization, the flattening behavior might be interpreted as a signal of chiral-symmetry restoration.

It is informative to refer to results obtained using other fermion formulations. Here, we mention a study using rooted staggered fermions, where the chiral symmetry is better preserved~\cite{Astrakhantsev:2020tdl}.
%%%%%%%%%%%%%%%%%%%%%%%%%%%%%%%%%%%%%%%%%%%%%%%%%%%%%%%%%%
\begin{figure}[h]
\centering
\includegraphics[width=0.8 \textwidth]{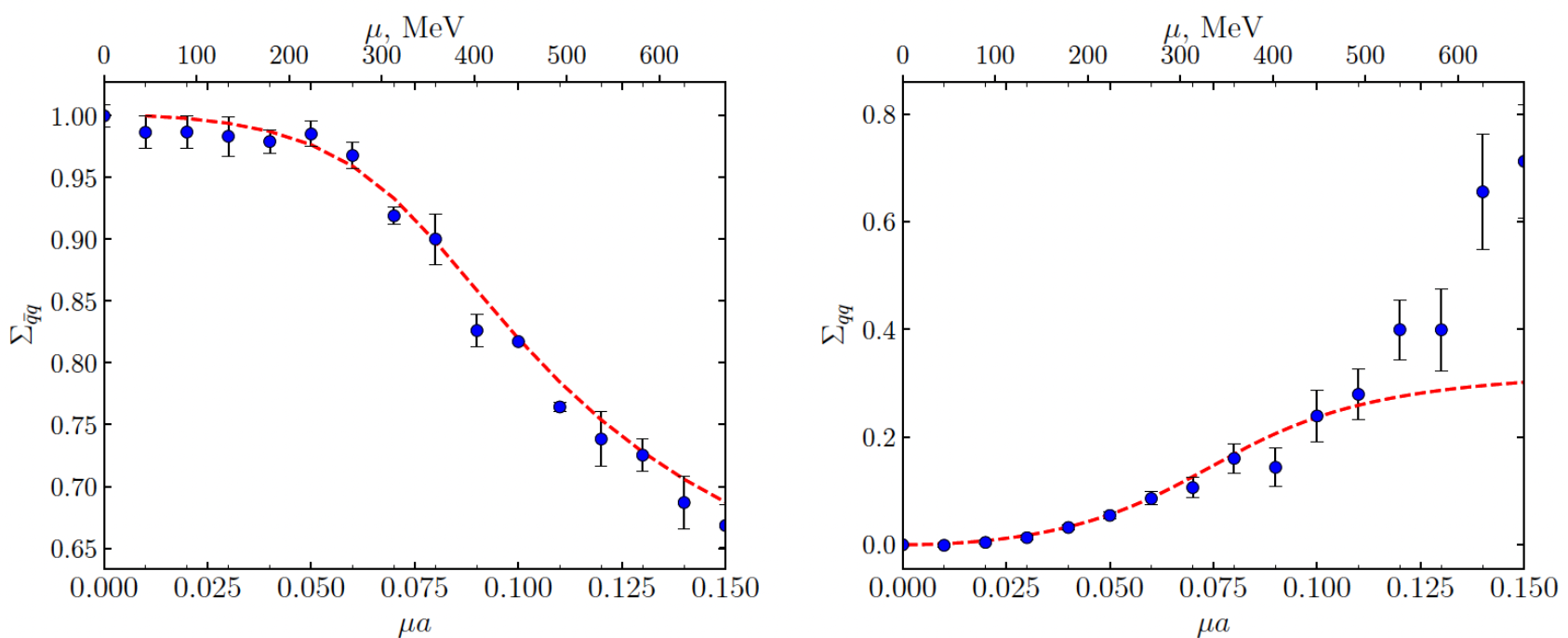}
\caption{Renormalized chiral condensates (left) and diquark condensate (right) as a function of $a\mu$ using the staggered fermions. It is taken from Ref.~\cite{Astrakhantsev:2020tdl} } \label{fig:chiral-diquark-staggered}
\end{figure}
%%%%%%%%%%%%%%%%%%%%%%%%%%%%%%%%%%%%%%%%%%%%%%%%%%%%%%%%%%
Figure~\ref{fig:chiral-diquark-staggered} shows the results for the renormalized chiral condensate (left panel) and the renormalized diquark condensate (right panel) obtained by the simulations with rooted staggered fermions. Here, the renormalization is performed by subtracting the corresponding $\mu=0$ values:
\begin{align}
& \Sigma_{\bar{q}q} = \frac{m}{4m_\pi^2F^2} \left[ \langle \bar{q} q \rangle (\mu) - \langle \bar{q} q \rangle (0) \right] +1, \nonumber\\
& \Sigma_{qq} = \frac{m}{4m_\pi^2F^2} \left[  \langle qq \rangle (\mu) - \langle qq \rangle (0)\right] .\label{eq:renorm-qq}
\end{align}

In this analysis, the diquark source parameter (denoted as $\lambda$ in this work) is kept finite. The leading-order ChPT expressions are modified to account for finite quark mass $m$ and finite $\lambda$, combined into a correction term defined as
\begin{align}
& \langle \bar{q}q \rangle = 2N_f G \cos \alpha, \quad \langle qq \rangle = 2N_f G \sin \alpha,
\label{eq:alpha_definition}    
\end{align}
where $\alpha$ is determined by 
\begin{align}
 \mu^2 \sin \alpha \cos \alpha = \mu_c^2 \left( \sin \alpha -\frac{\lambda}{m} \cos \alpha \right).
\end{align}
Here, the constant $G$ is given by
\beq
G=\frac{1}{N_f} \sqrt{\langle \bar{q}q \rangle^2 + \langle qq \rangle^2}.
\eeq
This relation, where the squares of the chiral and diquark condensates add up to a constant $G$, reflects a remnant of the extended flavor symmetry, namely the Pauli-G\"{u}rsey symmetry, that exists at \( m = \mu = 0 \). It has also been investigated in Ref.~\cite{Kogut:1999iv, Kogut:2001na} previously.

The fitting function, Eq.~\eqref{eq:renorm-qq}, with this correction and a universal parameter $F$, which is related to the pion decay constant by $f_\pi = F/2$, is used for both the chiral and diquark condensates. It is found that in the BEC region ($a\mu \in (0, 0.12)$ in this work), both condensates can be well described by the same value of $F$, supporting the validity of the ChPT description in this regime.
%%%%%%%%%%%%%%%%%%%%%%%%%%%%%%%%%%%%%%%%%%%5

%%%%%%%%%%%%%%%%%%%%%%%%%%%%%%%%%%%%%%%%%%%%%
\subsection{Confinement or deconfinement in high density regime}\label{sec:confinement}
%%%%%%%%%%%%%%%%%%%%%%%%%%%%%%%%%%%%%%%%%%%%%
In recent years, there has been interest in clarifying whether the BCS phase at high density is confined or deconfined.
Conventionally, the Polyakov loop has been used as a simple order parameter for confinement, and early studies employed it to explore the possible emergence of deconfinement at high densities. Since the high-density regime is expected to be governed by perturbative QCD, it was natural to anticipate deconfinement in this limit. For instance, a pioneering study by S.~Hands et al.~\cite{Hands:2006ve} reported a gradual increase in the Polyakov loop at $T=45$ MeV as the density increased, which was interpreted as an indication of deconfinement.

\begin{figure}[h]
\centering
\includegraphics[width=0.9 \textwidth]{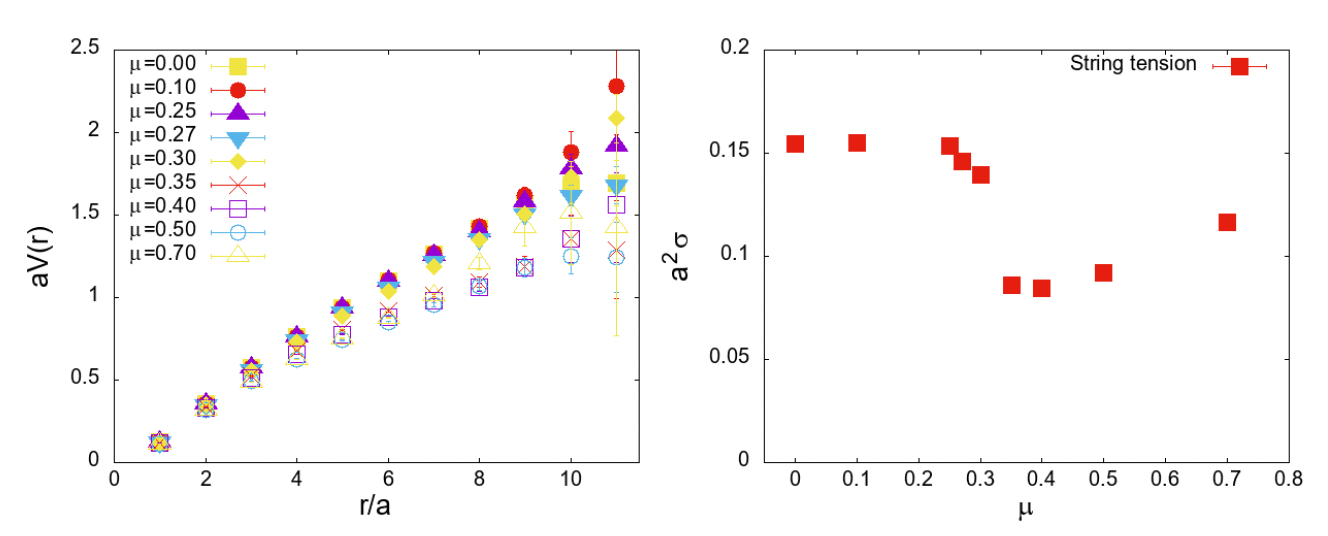}
\caption{The $\mu$ dependence of the $q-\bar{q}$ potential and string tension. The plots originally appeared in Ref.~\cite{Ishiguro:2021yxr}. } \label{fig:string-tension}
\end{figure}

However, in our previous study~\cite{Iida:2019rah}, we observed strong lattice artifacts that appeared before the onset of the expected deconfinement transition at $T=80$ MeV. These results suggest that confinement persists throughout the BCS phase at this temperature.
To obtain a more reliable probe of confinement, we investigated the static quark–antiquark potential~\cite{Ishiguro:2021yxr}. Our analysis shows a linear increase with separation even at the highest density explored at $T=40$ MeV, providing strong evidence that the confining force remains intact in this regime.
Thus, we investigate the $\mu$-dependence of the string tension by fitting these potentials using the Cornell-type function,
\beq
V(r)=\sigma r + \frac{c}{r} + V_0. 
\eeq
The result of string tension is almost constant in the hadronic phase, while it decreases around but still has a non-zero value even at the highest $\mu$ as shown in Figure~\ref{fig:string-tension}.
 %%%%%%%%%%%
%%%%%%%%%%%%%%%%%%%%%%%%%%%%%%%%%%%%%%%%%%%%%%%%%%%%%%%%%%

\begin{figure}[h]
\centering
\includegraphics[width=0.6 \textwidth]{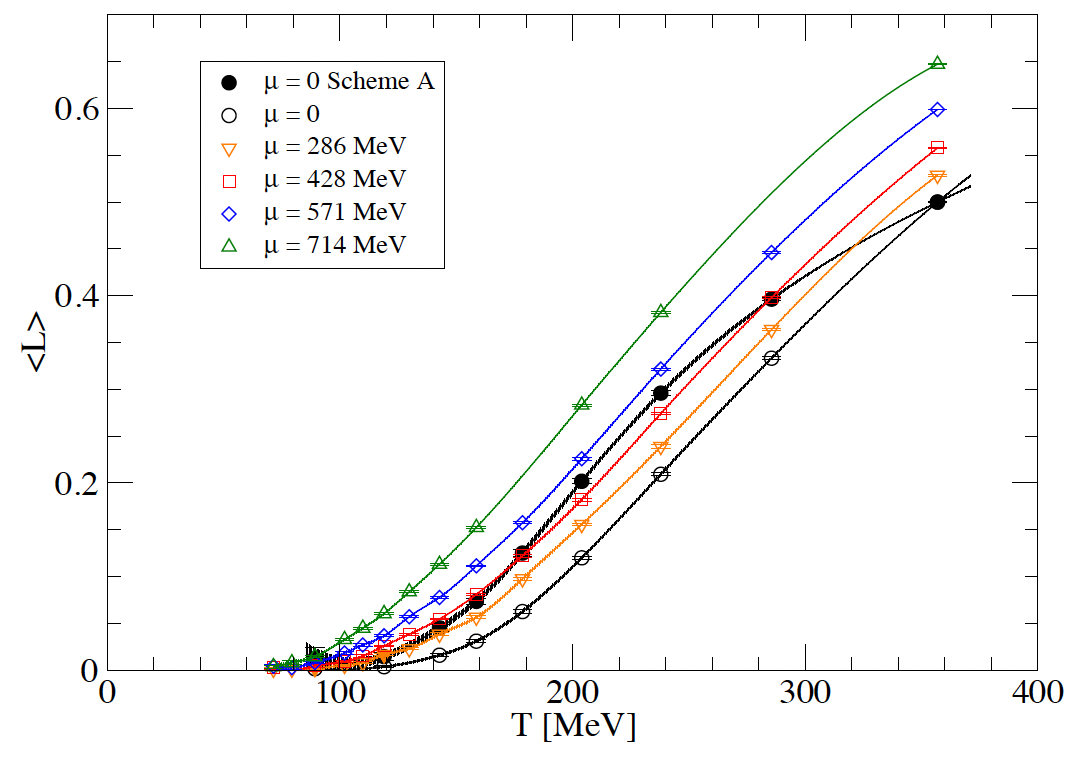}
\caption{Renormalized Polyakov loop as a function of temperature. It is taken from Ref.~\cite{Boz:2019enj}. } \label{fig:Polyakov-Hands}
\end{figure}

Soon after Ref.~\cite{Iida:2019rah}, S.~Hands and collaborators revisited their earlier analysis with simulations on finer lattices and larger volumes~\cite{Boz:2019enj}. By employing two different renormalization schemes, they performed a detailed analysis of the Polyakov loop behavior. Their updated results confirmed that the renormalized Polyakov loop remains close to zero even at $\mu \approx 700$ MeV, as long as the temperature is below approximately $T=100$ MeV, reinforcing the conclusion that confinement can persist deep into the high-density regime (see Figure~\ref{fig:Polyakov-Hands}).
%%%%%%%%%%%
%%%%%%%%%%%%%%%%%%%%%%%%%%%%%%%%%%%%%%%%%%%%%%%%%%%%%%%%%%
It is found that the Polyakov loop at high density is highly sensitive to finite-volume effects.

%%%%%%%%%%%%%%%%%%%%%%%%%%%%%%%%%%%%%%%%%%%%%%%%%%%%%%%%%%
\begin{figure}[h]
\centering
\includegraphics[width=0.45 \textwidth]{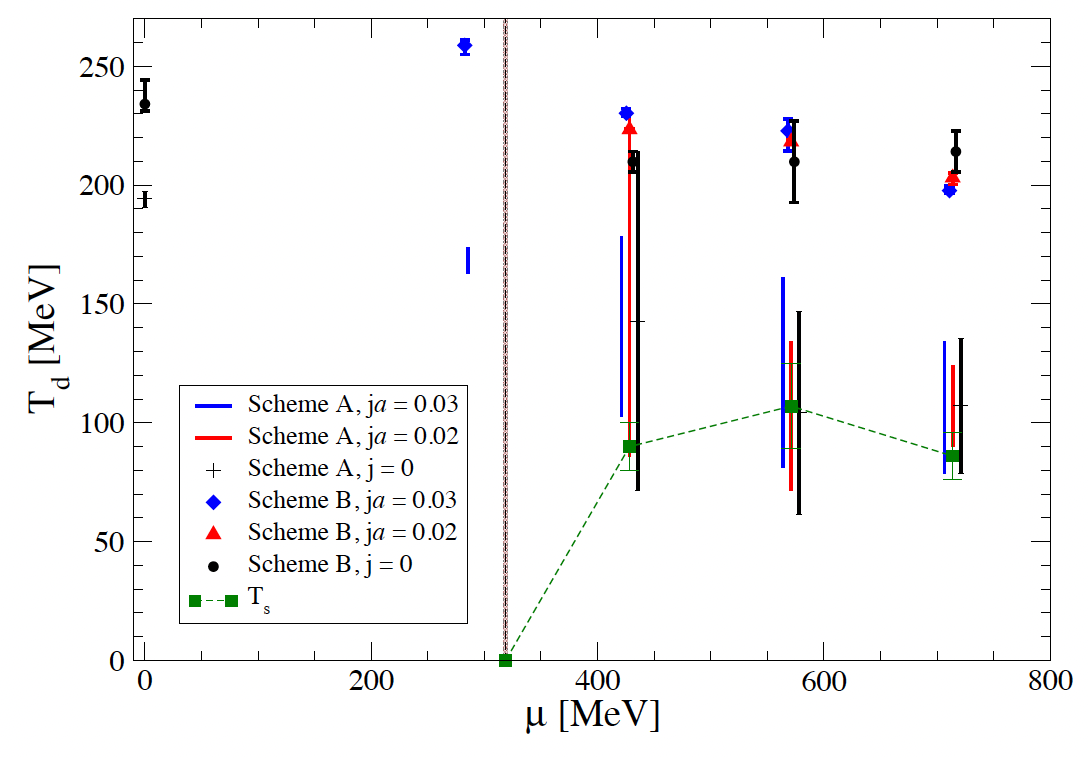}
\includegraphics[width=0.45 \textwidth]{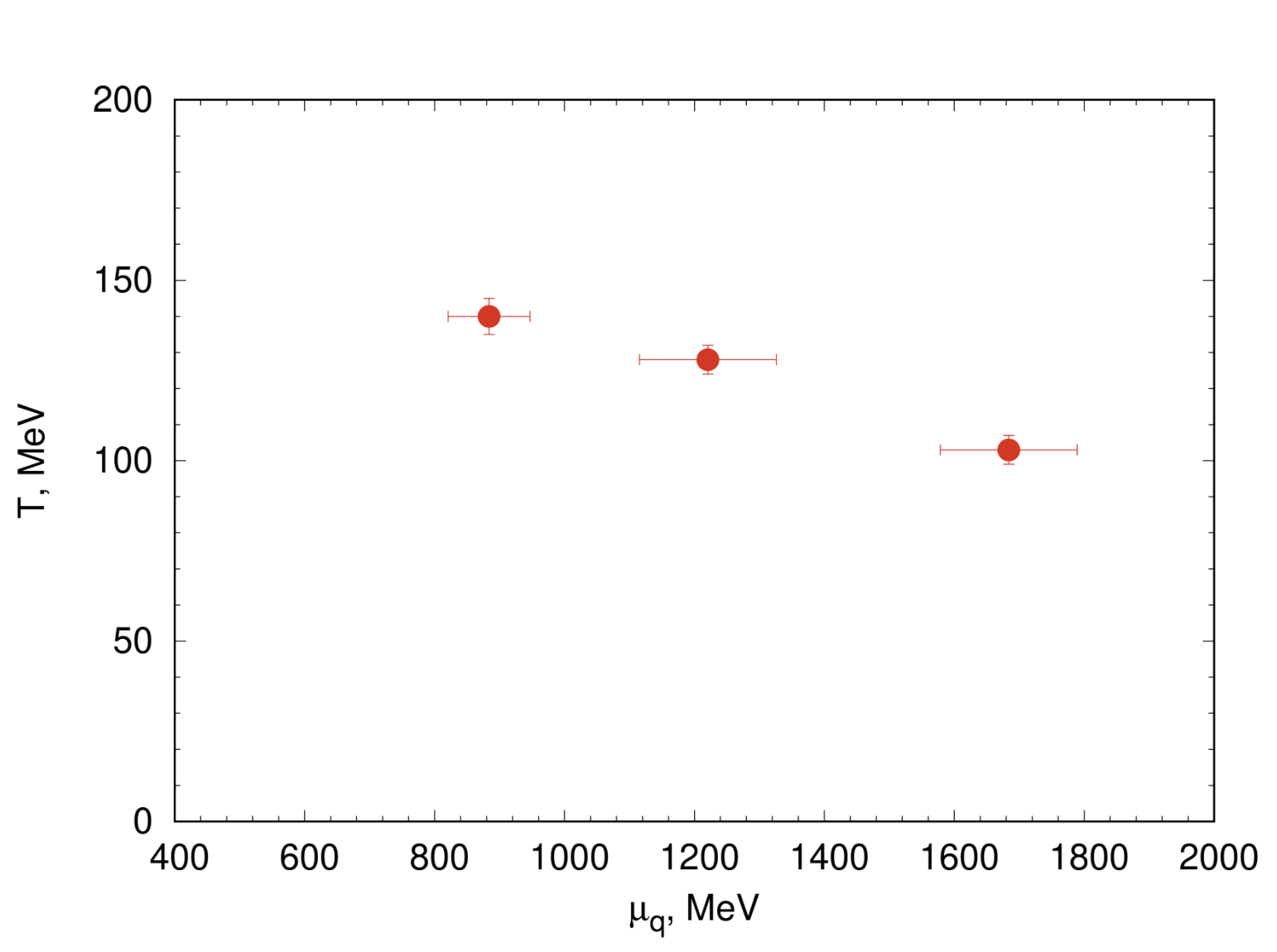}
\caption{The confinement-deconfinement transition temperature ($T_d$) in ($\mu$--T) plane. The result using Wilson fermion (left) and staggered fermion (right), originally appeared in Ref.~\cite{Boz:2019enj} and ~\cite{Begun:2022bxj}, respectively. } \label{fig:conf-deconf}
\end{figure}
%%%%%%%%%%%%%%%%%%%%%%%%%%%%%%%%%%%%%%%%%%%%%%%%%%%%%%%%%%

In the same paper~\cite{Boz:2019enj}, by comparing the deconfinement temperature $T_d$ extracted from the renormalized Polyakov loop with the superfluid transition temperature $T_s$, the authors found intriguing behavior (see the left panel of Figure~\ref{fig:conf-deconf}). Specifically, the results suggest that $T_d$ is slightly above or nearly equal to the superfluid transition temperature $T_s \simeq 90$~MeV. If this behavior persists in the continuum limit, it implies that no deconfinement transition occurs at $T=0$, and that the entire superfluid phase remains confined for all values of chemical potential.
Furthermore, another group, using the rooted staggered fermions~\cite{Begun:2022bxj}, also investigated the critical temperature for deconfinement, $T_d$, in a high-density regime; $T_d \approx 100$~MeV was found even at $\mu \approx 1.8$~GeV (see the right panel of Figure~\ref{fig:conf-deconf}). This result also suggests that the deconfinement temperature becomes nearly independent of $\mu$ at large chemical potentials, indicating that confinement persists at low temperatures even in high-density regimes.

Based on these considerations, the QC$_2$D system remains confined in the BCS phase at low temperature, as shown in Figure~\ref{fig:comp-phase-diagram}~\footnote{ A theoretical analysis based on ’t Hooft anomaly matching in massless QC$_2$D yields the constraint $T_d \le T_s$ at fixed $\mu$~\cite{Furusawa:2020qdz}. 
In Figure~\ref{fig:comp-phase-diagram}, we assume $T_d < T_s$; following this analysis is valid in the massive fermion case.}

%%%%%%%%%%%%%%%%%%%%%%%%%%%%%%%%%%%%%%%%%%%%%
\subsection{Topological susceptibility}
%%%%%%%%%%%%%%%%%%%%%%%%%%%%%%%%%%%%%%%%%%%%%
At low temperatures, the system undergoes a sequence of phase transitions as the chemical potential increases; from the hadronic phase to the BEC phase, and eventually to the BCS-like superfluid phase. To further characterize these phases, we now turn to the behavior of (semi-)classical configurations in each regime.

One particularly interesting nonperturbative object is the topological charge distribution, characterized by the topological susceptibility, $\chi_Q$, given in Eq~\eqref{eq:def-chi_Q}. 
A pioneering study investigated this quantity in the case of $N_f=4$ and $8$ staggered fermions~\cite{Hands:2011hd, Alles:2006ea}. Their results showed that $\chi_Q$ decreases in the high-density regime. Indeed, the figure demonstrates a correlated increase in the Polyakov loop with a suppression of topological fluctuations.

%%%%%%%%%%%%%%%%%%%%%%%%%%%%%%%%%%%%%%%%%%%%%%%%%%%%%%%%%%
\begin{figure}[htbp]
\centering
\includegraphics[width=1.0 \textwidth]{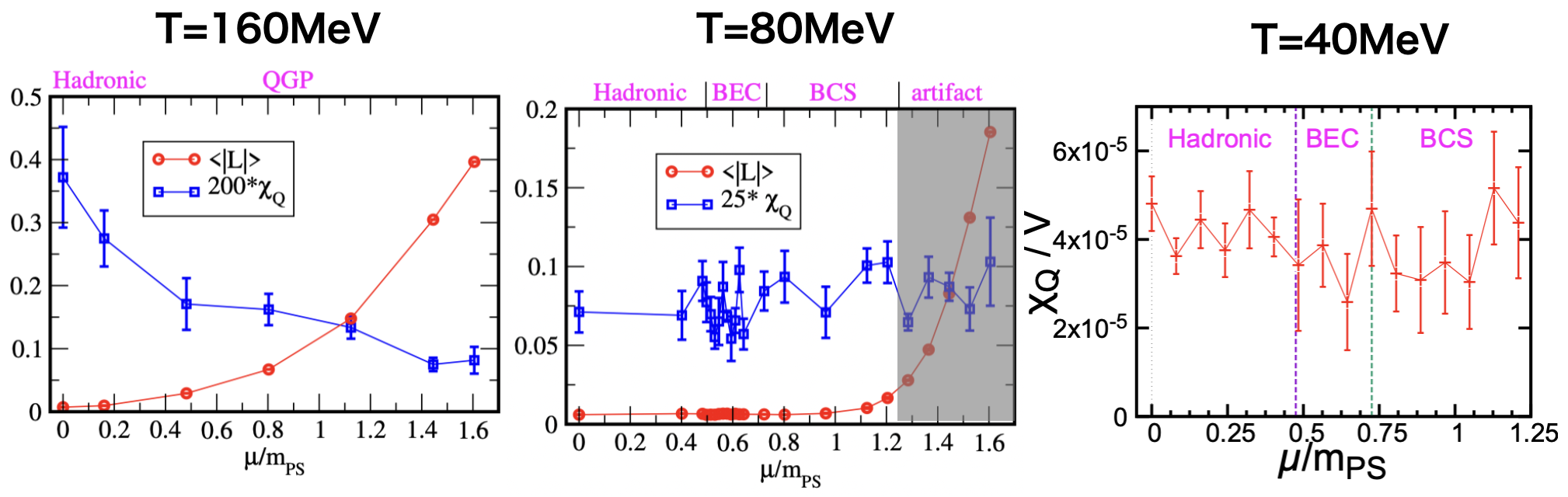}
\caption{The $\mu$-dependence of the topological susceptibility at $T = 160$ MeV, 80 MeV (blue data), and 40 MeV (red data). At $T = 160$ MeV and 80 MeV, we also show the magnitude of the Polyakov loop (red data) to see the confining behavior. The data are originally obtained in Refs.~\cite{Iida:2019rah, Iida:2024irv}. }\label{fig:chi-Q-old}
\end{figure}
%%%%%%%%%%%%%%%%%%%%%%%%%%%%%%%%%%%%%%%%%%%%%%%%%%%%%%%%%%
However, our recent studies across various temperatures revealed a more nuanced behavior. At $T = 160$ MeV on $32^3 \times 8$ lattices, where the hadronic–QGP transition occurs, $\chi_Q$ indeed decreases with increasing chemical potential, consistent with the previous findings. In contrast, at lower temperatures, at $T = 80$ MeV on $16^4$ lattices and $T = 40$ MeV on $32^4$ lattices, where confinement persists, we found that $\chi_Q$ remains approximately constant as a function of $\mu$ as shown in Figure~\ref{fig:chi-Q-old}. This behavior suggests that nontrivial topological configurations remain unsuppressed in the confined, dense regime.
To our knowledge, this persistence of topological susceptibility at low temperature and high density has not yet been confirmed by other lattice groups, and it remains an open question~\footnote{In Ref.~\cite{Astrakhantsev:2020tdl}, which employed the rooted staggered fermions, it was shown that the topological susceptibility $\chi_Q$ decreases in the high-density regime. However, it should be noted that this simulation was performed at a relatively high temperature, around $T \sim 140$~MeV, where the system is already known to be in the deconfined phase. Therefore, the suppression of $\chi_Q$ observed in that study is likely associated with deconfinement rather than being a universal feature of dense QC$_2$D.
}.

In summary, we have seen that quark-based observables, such as the diquark condensate and quark number density, are well described by weak-coupling analyses and free-quark models in the high-density BCS phase. The behavior of the chiral condensate also indicates a gradual restoration of chiral symmetry as the chemical potential increases.
In contrast, gluonic observables, including the topological susceptibility, the Polyakov loop, and the static quark–antiquark potential, exhibit behavior that remains nonperturbative properties at $\mu = 0$ and low temperature. This suggests that, even at $T=0$ and $\mu \to \infty$, the system retains characteristics of confinement and topological effects.
Taken together, these results imply that the dense and cold region of QC$_2$D realizes a novel nonperturbative regime that is distinct from both the hadronic phase at low temperatures and densities, and the QGP phase at high temperatures.

%%%%%%%%%%%%%%%%%%%%%%%%%%%%%%%%%%%%%%%%%%
\section{Equation of state and sound velocity}\label{sec:EoS}
%%%%%%%%%%%%%%%%%%%%%%%%%%%%%%%%%%%%%%%%%%%%%
\subsection{Overview}
In this section, we discuss the equation of state (EoS) and the speed of sound ($c_{\rm s}$) in dense QC$_2$D, focusing in particular on novel features that have emerged in recent years.

\begin{figure}[htbp]
\centering
\includegraphics[width=.8\textwidth]{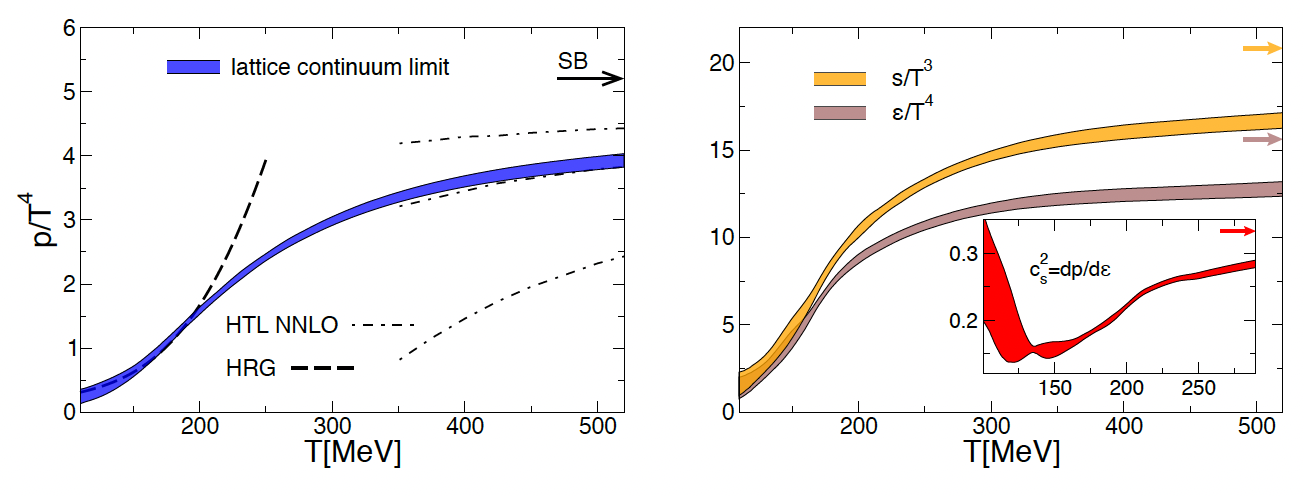}
\includegraphics[width=.8\textwidth]{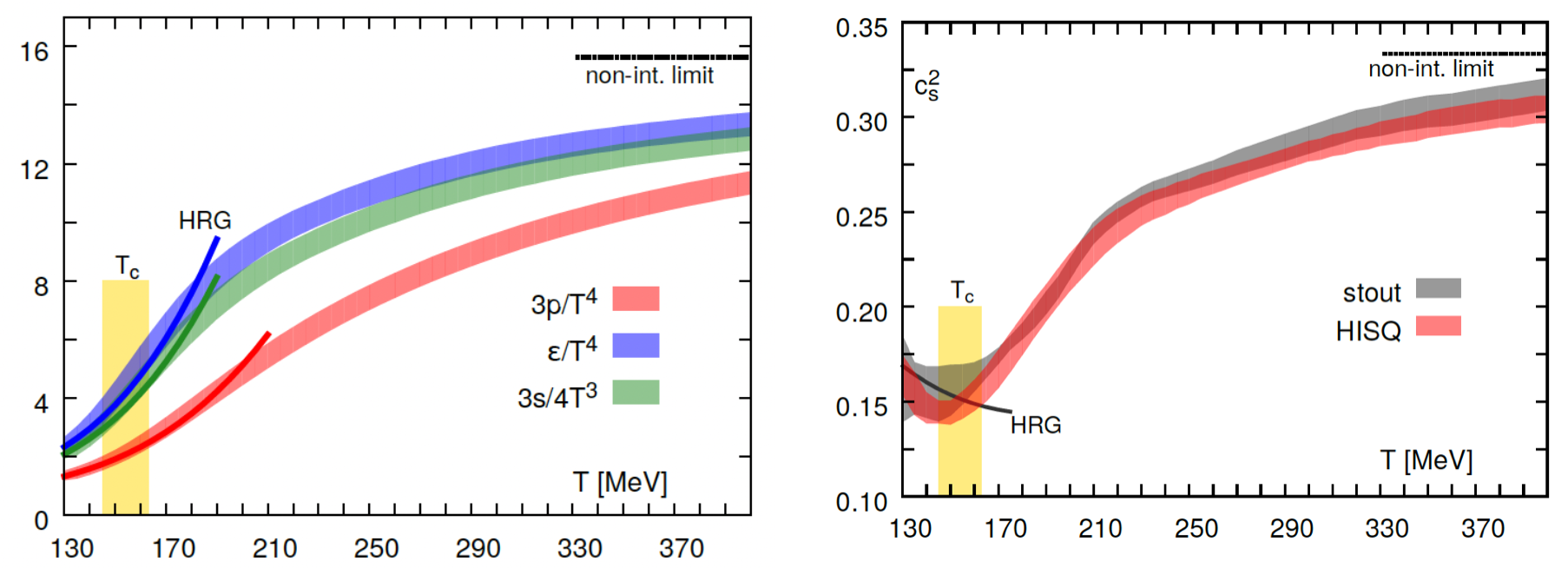}
\caption{Temperature dependence of thermodynamic quantities and the speed of sound ($c_{\rm s}$) at $\mu = 0$. The results were originally presented in Refs.~\cite{Borsanyi:2013bia} (top) and~\cite{HotQCD:2014kol} (bottom).}
\label{fig:finite-T}
\end{figure}

At zero chemical potential, the EoS and associated thermodynamic quantities at finite temperature have been extensively studied in the context of the early universe, where they serve as inputs for the matter content in the Einstein equations. Through Monte Carlo simulations, significant progress has been made in determining these quantities with high precision, ranging from pure Yang–Mills theory to full QCD with dynamical quarks.
For instance, in $N_f = 2+1$ QCD, the pressure, energy density, and entropy density show a rapid growth around the chiral (or pseudocritical) transition temperature ($T_c$) and then increase monotonically with temperature, approaching the Stefan–Boltzmann (SB) limit at high temperatures, as illustrated in Figure~\ref{fig:finite-T}. 
The speed of sound, defined through the thermodynamic relation $c_{\rm s}^2/c^2 = \partial p / \partial \epsilon$, shows a local minimum around $T_c$ and subsequently increases monotonically in the QGP phase toward the Stefan–Boltzmann (SB) limit, $c_{\rm s}^2/c^2 = 1/3$ (with $e = 3p$). This value represents an upper bound for the squared sound velocity, often referred to as the conformal bound.
%%%%%%%%%%%%%%%%%%%%%%%%%%%%%%%%%%%%%%%%%%%%%%%%%%%%%%%%%%
%%%%%%%%%%%%%%%%%%%%%%%%%%%%%%%%%%%%%%%%%%%%%%%%%%%%%%%%%%

However, recent lattice studies of dense QC$_2$D and also three-color QCD at isospin density have revealed qualitatively different behavior in the low-temperature and high-density regimes. Contrary to the conventional picture, the speed of sound in these systems does not increase monotonically, and can exceed the conformal bound.
The first indication of this phenomenon appeared in Ref.~\cite{Iida:2022hyy} in 2022. Since then, the same qualitative behavior has been confirmed by four independent groups using different setups and lattice actions~\cite{Brandt:2022hwy, Abbott:2023coj, Abbott:2024vhj, Iida:2024irv, Hands:2024nkx}. Here, we review recent developments on this topic, with a particular focus on our results originally presented in Refs.~\cite{Iida:2022hyy, Iida:2024irv}, which provides a comprehensive analysis of the dense regime.

\subsection{Calculation strategy}
To study EoS at finite density, it is essential to evaluate thermodynamic quantities such as the pressure $p(\mu)$ and the energy density $e(\mu)$ on the lattice. 
At finite chemical potential, the pressure, in the thermodynamic limit, satisfies the Gibbs-Duhem relation, which allows us to calculate the pressure as
\beq
p (\mu) = \int_{\mu_c}^\mu n_q(\mu') d \mu' , \label{eq:def-p}
\eeq
where $\mu_c$ denotes the critical value of $\mu$ for the hadronic-superfluid phase transition.
Note that there is no renormalization for the quark number density as it is a conserved quantity.
On the lattice, two schemes with different discretization errors have been proposed in Ref.~\cite{Hands:2006ve}.
Here, we show results defined by
\beq
\frac{p}{p_{SB}}(\mu) &= \int_{\mu_o}^{\mu} d\mu' \frac{n_{SB}^{cont.}}{n_{q}^{\mathrm{tree}}}   n^{latt.}_q(\mu')  / \int_{\mu_o}^{\mu} d\mu' n_{SB}^{cont.} (\mu'),\label{eq:p-scheme2}
\eeq
which given as Scheme~II in Ref.~\cite{Hands:2006ve}.
Here, $p_{SB}(\mu)$ denotes the pressure value of the non-interacting theory, namely the SB limit, which is obtained by the numerical integration of the number density of quarks in the relativistic limit.
In the continuum theory at zero-temperature, the pressure scales as $p_{SB}(\mu) = \int^\mu n_{SB}^{cont.}(\mu')d\mu' \approx N_fN_c \mu^4 /(12\pi^2)$ in the high $\mu$ regime.
Also, $\mu_o$ in Eq.~\eqref{eq:p-scheme2} represents the starting point at which $\langle n_q \rangle$ becomes non-zero as $\mu$ increases. In the actual calculation, we only have discrete values of $\mu$ as simulation parameters.
Furthermore, strictly speaking, $\mu_c \ne \mu_o$ and $\mu_c \ne  m_{\rm PS}/2$ in actual numerical data, because of the calculation with non-zero temperature. However, in our work for $T=40$ MeV, we found $a\mu_o = 0.30$ ($\mu/ m_{\rm PS}=0.48$), which is very close to $\mu_c$ obtained in Sec.~\ref{sec:hadron-SF}.

To obtain both $e(\mu)$ and $p(\mu)$, respectively, we need another observable. Here, we measure the trace anomaly, namely $e-3p$.
The trace anomaly basically consists of the beta-function for the parameters in the action and the trace part of the energy-momentum tensor;
\beq
e-3p &=& \frac{1}{N_s^3 N_\tau} \left( \left. a \frac{d \beta}{d a} \right|_{\mathrm{LCP}} \left\langle \frac{\partial S}{\partial \beta} \right\rangle_{sub.}  + \left. a \frac{d  \kappa}{d a} \right|_{\mathrm{LCP}} \left\langle \frac{\partial S}{\partial \kappa} \right\rangle_{sub.} + \left. a \frac{\partial j}{\partial a}\right|_{\mathrm{LCP}} \left\langle \frac{\partial S}{\partial j} \right\rangle_{sub.} \right). \nonumber \\  \label{eq:trace-anomaly}
\eeq
Here, there is no $\mu$-derivative term, since the associated observable, namely quark number, is the renormalization-conserved quantity.

How to estimate the beta-function or multiplicative renormalization factor at finite $\mu$ is nontrivial.
In the studies of dense QC$_2$D, several methods have been performed, such as non-perturbative determination of Karsch coefficients~\cite{Karsch:1989fu, Hands:2011ye}, the usage of the Wilson line operator~\cite{Hands:2024nkx}, and the perturbative calculations~\cite{Astrakhantsev:2020tdl}.
In our works~\cite{Iida:2022hyy, Iida:2024irv}, we utilize a non-perturbative beta-function for each parameter, which can be evaluated at $\mu=0$ using the $w_0$ scale proposed in Ref.~\cite{BMW:2012hcm} and a fixed mass-ratio between the pseudo-scalar and vector mesons, namely $m_{\rm PS}/m_V$, along the line of constant physics (LCP).
According to the data in Ref.~\cite{Iida:2020emi}, we obtained
\beq
 a d\beta /da|_{\beta=0.80,\kappa=0.159}=-0.352, \quad a d\kappa/da |_{\beta=0.80,\kappa=0.159}=0.0282.\label{eq:beta-fn}
\eeq 

In Eq.~\eqref{eq:trace-anomaly}, we take all observables, $\langle \mathcal{O} \rangle$, in the $j \rightarrow 0$ limit. 
More specifically, $\langle \mathcal{O} \rangle_{sub.} (\mu) $ denotes the quantity that has the vacuum contribution subtracted out. 
Ideally, we take $\langle \mathcal{O} \rangle_{sub.} (\mu) = \langle \mathcal{O} (\mu,T)  \rangle - \langle \mathcal{O} (\mu=0,T=0) \rangle $, but the simulation at absolute zero temperature is practically difficult.
Therefore, we perform a fixed $T$  subtraction to see the $\mu$-dependence, namely,  $\langle \mathcal{O} \rangle_{sub.} (\mu) = \langle \mathcal{O} (\mu, T\approx 40 \mathrm{~MeV})  \rangle - \langle \mathcal{O} (\mu=0, T \approx 40 \mathrm{~MeV}) \rangle $. 
In fact, we invested the temperature dependence of $\langle \mathcal{O} (\mu=0,T) \rangle$ below $T_c$, then we found that it is negligible~\cite{Iida:2024irv}.

The first term of the RHS in Eq.~\eqref{eq:trace-anomaly}, $\left\langle \partial S / \partial \beta \right\rangle_{sub.} $, is given by the measurement of the gauge action, where we evaluate the value of the Iwasaki gauge action for generated configurations. 
The second term essentially expresses the fermion contributions to the trace anomaly,  which can be evaluated by
\beq
\left\langle \frac{\partial S}{\partial \kappa} \right\rangle = \frac{1}{\kappa} \left( \mathrm{Tr}_{c, s,f}\mathbb{I} - N_f \langle \bar{q}q \rangle \right).\label{eq:trace-anomaly-fermion}
\eeq
Thus, we also measure the chiral condensate, which has already been investigated in Figure~\ref{fig:scaling-chiral-cond}.

Finally, we evaluate the sound velocity using a symmetric difference to obtain $\partial p /\partial \mu$ and $\partial e /\partial \mu$;
\beq
c_\mathrm{s}^2 (\mu)/c^2= \frac{\Delta p (\mu)}{\Delta e (\mu)} = \frac{ p(\mu +\Delta \mu) - p(\mu -\Delta \mu)}{e(\mu +\Delta \mu) - e(\mu -\Delta \mu)}\label{eq:sound-velocity}
\eeq
at a fixed temperature.
Note that in the standard definition, the sound velocity squared is given by $\partial p /\partial e |_{s=\mathrm{const.}}$ where $s$ denotes the entropy per baryon, but here we calculate $\partial p /\partial e |_{T=\mathrm{const.}}$.
It is technically hard to evaluate the former one in lattice simulations. 
At $T=0$, both quantities are equivalent to each other. 
Careful study of the temperature dependence of the latter one is an important task and thus will be addressed below.

\subsection{Results: Thermodynamic quantities and decay constant}
\begin{figure}[htbp]
\centering
\includegraphics[width=.4\textwidth]{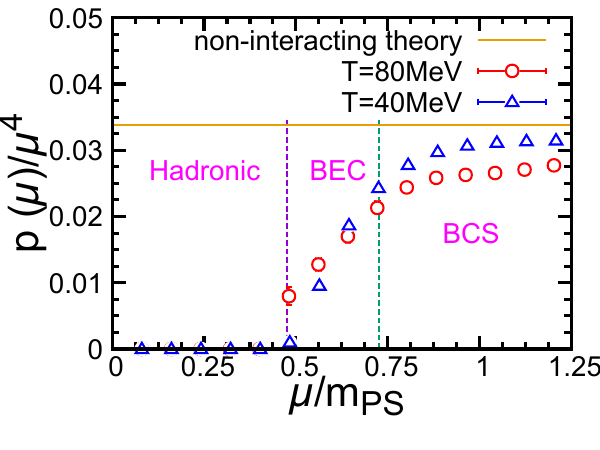}
\qquad
\includegraphics[width=.4\textwidth]{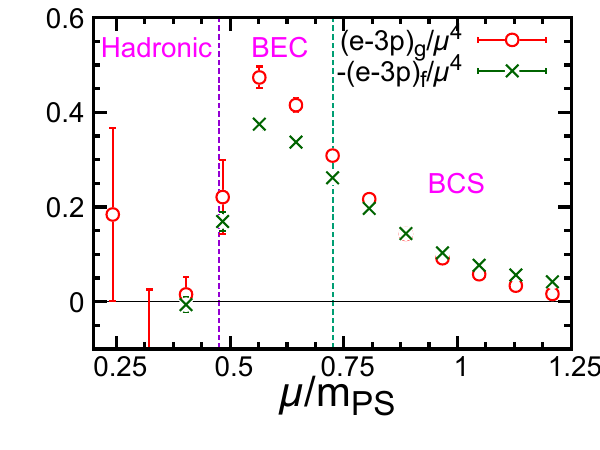}
\caption{Pressure (left panel) and trace anomaly (right panel) as a function of $\mu$.   The right panel depicts the first (circle-red symbol) and second (cross-green symbol) terms in Eq.~\eqref{eq:trace-anomaly} at $T = 40$ MeV separately. The plots are originally shown in Ref.~\cite{Iida:2024irv}.}\label{fig:p-Tdeps}
\end{figure}

The results for the pressure are shown in the left panel of Figure~\ref{fig:p-Tdeps}. In the hadronic phase, where $\langle n_q \rangle $ is consistent with zero, the pressure is also consistent with zero. Once the superfluid phase transition occurs, the pressure increases monotonically. At $T = 40$ MeV (triangle-blue symbols), the pressure grows sharply in the BEC phase and approaches the SB limit (orange line) more closely in the high-density region. 

%%%%%%%%%%%%%%%%%%%%%%%%%%%%%%%%%%%%%%%%%%%%%%%%%%%%%%%%%%
%%%%%%%%%%%%%%%%%%%%%%%%%%%%%%%%%%%%%%%%%%%%%%%%%%%%%%%%%%
On the other hand, we plotted the first (gluonic) term and minus the second (fermionic) term of the trace anomaly~\eqref{eq:trace-anomaly} at $T = 40$ MeV as $\langle e- 3p\rangle_g$ (circle-red symbols) and  $- \langle e -3p \rangle_f$ (cross-green symbols), respectively, in the right panel of Figure~\ref{fig:p-Tdeps}. Note that the second term takes negative values, which have the sign flipped in the plot. Furthermore, we simply neglected the third term in Eq.~\eqref{eq:trace-anomaly} in our analysis. Although a careful discussion is required, we assumed here that no \( j \)-dependence remains in the \( j \to 0 \) limit.
Thus, the total trace anomaly is given by the circle-red data minus the cross-green data.
As can be seen from this plot, the trace anomaly is also zero in the hadronic phase. After the superfluid phase transition, the trace anomaly reaches a maximum value and then decreases. Notably, in the middle of the BCS phase, $- \langle e - 3p \rangle_f$ becomes larger than $\langle e -3p\rangle_g$, causing the trace anomaly to change from positive to negative.  
Indeed, such a negative trace anomaly in medium has also been predicted in the context of some effective models for dense QC$_2$D and 3-color QCD with isospin chemical potential~\cite{Lu:2019diy, Kawaguchi:2024iaw}. Furthermore, the trace anomaly at $\mu=T=0$ itself has been predicted to be negative owing to the structure of the QCD vacuum~\cite{Fukuda:1980py}.

\begin{figure}[htbp]
\centering
\includegraphics[width=.4\textwidth]{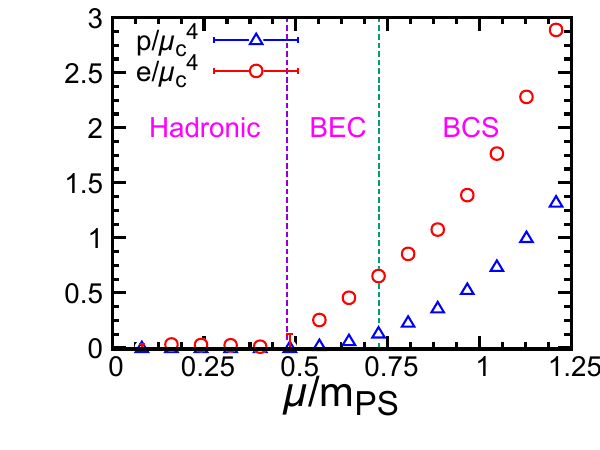}
\qquad
\includegraphics[width=.4\textwidth]{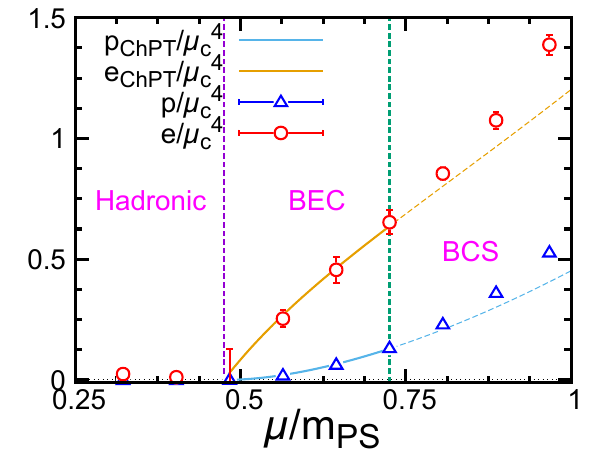}
\caption{(Left) The internal energy and pressure as a function of $\mu$. (Right) The enlarged plot of $p/\mu_c^4$ and $e/\mu_c^4$ around the BEC phase. The cyan and orange curves represent the fitting functions for $p/\mu_c^4$ and $e/\mu_c^4$, respectively, whose forms are given by 
the ChPT theory as shown in Eqs.~\eqref{eq:ChPT-p} and~\eqref{eq:ChPT-e}.  The plots are originally shown in Ref.~\cite{Iida:2024irv}.}\label{fig:e-and-p}
\end{figure}

Combining the results for the trace anomaly and pressure, we obtain the energy density and pressure as a function of $\mu/m_{\rm PS}$  as shown in the left panel of Figure~\ref{fig:e-and-p}.
Here, we normalize $e$ and $p$ by using $\mu_c^4$ in such a way as to be dimensionless.
As an initial consistency check, we confirm that $e$ and $p$ are consistent with zero in the hadronic phase, as expected.
%%%%%%%%%%%%%%%%%%%%%%%%%%%%%%%%%%%%%%%%%%%%%%%%%%%%%%%%%%
%%%%%%%%%%%%%%%%%%%%%%%%%%%%%%%%%%%%%%%%%%%%%%%%%%%%%%%%%%

At the end of this subsection, let us focus on the scaling behavior of $e$ and $p$ in the BEC phase.
In section~\ref{sec:phase-diagram}, we have shown that our results for the chiral condensate and diquark condensates are consistent with the predictions from ChPT.
It would therefore be worthwhile to fit the data for $p$ and $e$ in this regimes to the results of ChPT~\cite{Hands:2006ve},
\beq
p_{\rm ChPT}&=&4N_f F^2 \mu^2 \left( 1- \frac{\mu_c^2}{\mu^2} \right)^2, \label{eq:ChPT-p} \\
e_{\rm ChPT}&=&4N_f F^2 \mu^2 \left( 1- \frac{\mu_c^2}{\mu^2} \right) \left( 1+3 \frac{\mu_c^2}{\mu^2} \right), \label{eq:ChPT-e} 
\eeq
to obtain the pion decay constant ($F=f_\pi/2$) in QC$_2$D.

Figure~\ref{fig:e-and-p} depicts the $\mu$ dependence of $p$ and $e$ both for the data and fitted results. 
The good agreement of the data with the fit also ensures that our calculation of the beta-function~\eqref{eq:beta-fn} is reliable.
The best fit values  were obtained as $F= 51.1(5)$ MeV and $F=56.7(7)$ MeV from the fits of $p/\mu_c^4$ and $e/\mu_c^4$, respectively.
These values are similar to the earlier result, $F=60.8(1.6)$ MeV, obtained by independent simulation using the rooted staggered fermions in Ref.~\cite{Astrakhantsev:2020tdl}. 
Here, the authors performed the fit of the lattice data for the quark number density and the mixing angle between the diquark and chiral condensates around the phase transition point predicted by the ChPT analysis, as shown in Figure~\ref{fig:chiral-diquark-staggered}.

%%%%%%%%%%%%%%%%%%%%%%%%%%%%%%
\subsection{Results: Speed of sound}
\begin{figure}[htbp]
\centering
\includegraphics[width=0.6 \textwidth]{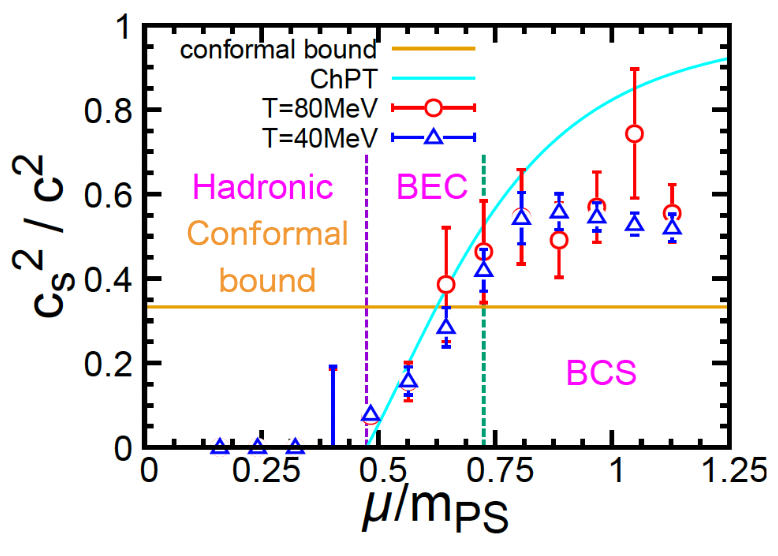}
\caption{The squared sound velocity at $T = 40$ MeV and $T = 80$ MeV. The cyan curve is the prediction of ChPT given in Eq.~\eqref{eq:cs-ChPT}. The horizontal line (orange) depicts the conformal bound, $c_{\rm s}^2/c^2 = 1/3$.  The plots are originally shown in Ref.~\cite{Iida:2024irv}.}\label{fig:sound-v}
\end{figure}

%%%%%%%%%%%%%%%%%%%%%%%%%%%%%%
We plotted the results for the speed of sound at $T = 40$ MeV and $T = 80$ MeV in Figure~\ref{fig:sound-v}.
%%%%%%%%%%%%%%%%%%%%%%%%%%%%%%%%%%%%%%%%%%%%%%%%%%%%%%%%%%
%%%%%%%%%%%%%%%%%%%%%%%%%%%%%%%%%%%%%%%%%%%%%%%%%%%%%%%%%%
Here, the cyan-curve expresses the prediction of the ChPT~\cite{Son_2001, Hands:2006ve},  
\beq
c_{\rm s}^2/c^2=(1-\mu_c^4/\mu^4)/(1+3\mu_c^4/\mu^4),\label{eq:cs-ChPT}
\eeq
where $c_{\mathrm s}^2/c^2 $ approaches $1$ in the high-density limit.  Our data are again consistent with this prediction in the BEC phase.  In the high-density region, namely the BCS phase, however, the values become smaller than the prediction of the ChPT analysis.
However, even in this BCS regime, our data exceed the conformal bound (shown as an orange line), which is known as $c_{\mathrm{s}}^2/c^2 = 1/3$.
By comparison of the results between T=80MeV and 40MeV, we found that the thermal effects are negligibly small, suggesting that the difference between the definitions of $\partial p/\partial e |_{s=const.}$ and $\partial p/\partial e |_{T=const.}$ as discussed below in Eq.~\eqref{eq:sound-velocity} is also negligibly small.
We can conclude that the excess over the conformal bound in dense QC$_2$D occurs at sufficiently low temperature, though taking the continuum and thermodynamic limits remains as future work.

Note that the pressure itself does not exceed the free-theory limit as shown in Figure~\ref{fig:p-Tdeps}. On the other hand, the pressure growth against the energy growth, corresponding to the sound velocity, is higher than the one for the free-theory,  which supports a stiff picture for QCD(-like) matter in the superfluid regime.

%%%%%%%%%%%%%%%%%%%%%%%%%%%%%%%%%%%%%%%%%%%%%%%%%%%%%%%%%%
\begin{figure}[htbp]
\centering
\includegraphics[width=0.6 \textwidth]{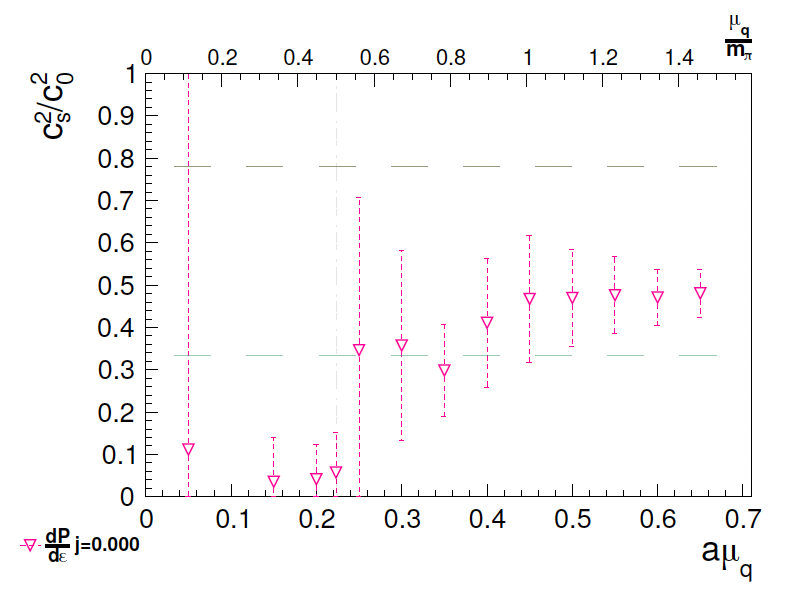}
\caption{Results of squared sound velocity for QC$_2$D given by S.~Hands et al. in Ref.~\cite{Hands:2024nkx}.}\label{fig:cs-Hands}
\end{figure}
%%%%%%%%%%%%%%%%%%%%%%%%%%%%%%%%%%%%%%%%%%%%%%%%%%%%%%%%%%
Finally, let us briefly introduce the results of the other calculations for dense QC$_2$D and three-color QCD with the isospin chemical potential.
A new preliminary result on the speed of sound in dense QC$_2$D was presented at the Lattice conference in 2024~\cite{Hands:2024nkx} (see Figure~\ref{fig:cs-Hands}). The result also indicates a violation of the conformal bound in the high-density regime.
Here, in Figure~\ref{fig:cs-Hands}, the gray dashed line represents a recently proposed upper bound on the speed of sound, $c_{\mathrm s}^2/c^2 \leq 0.781 $ derived from relativistic superfluid hydrodynamics~\cite{Hippert:2024hum}. Our results shown in Figure~\ref{fig:sound-v} also satisfy this new upper bound.

%%%%%%%%%%%%%%%%%%%%%%%%%%%%%%%%%%%%%%%%%%%%%%%%%%%%%%%%%%
\begin{figure}[htbp]
\centering
\includegraphics[width=0.9 \textwidth]{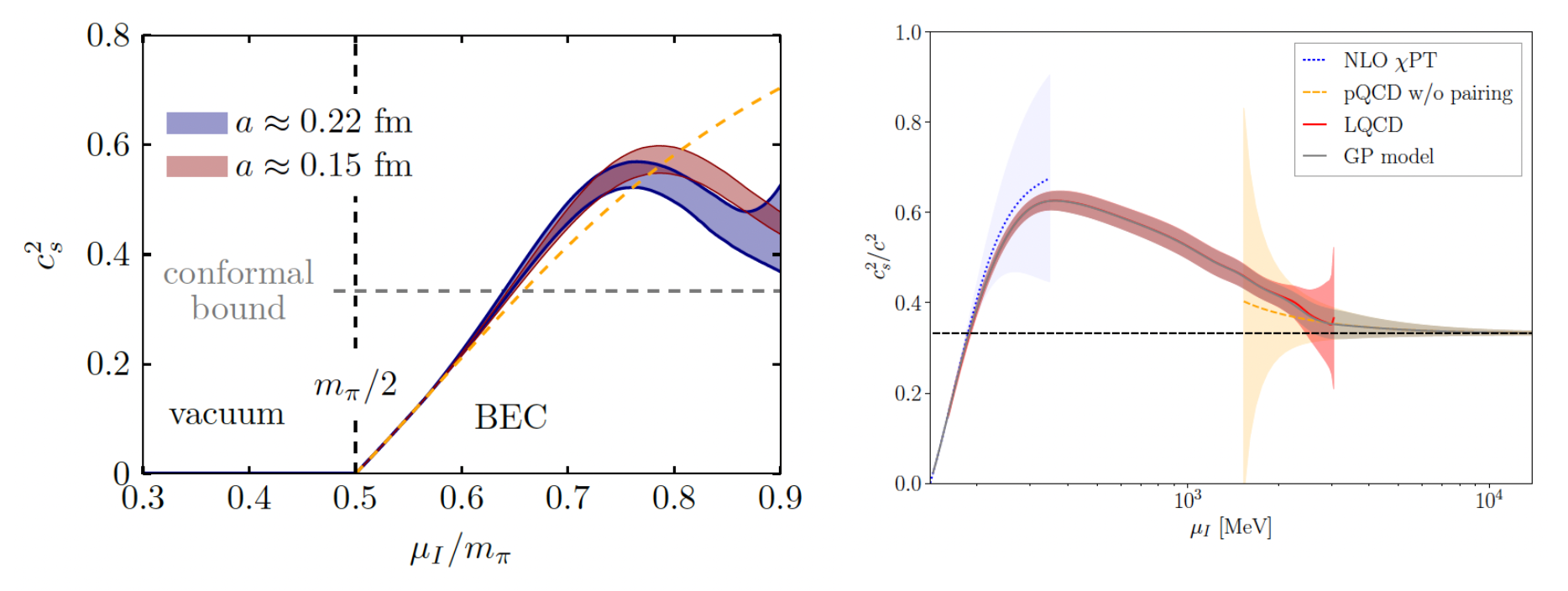}
\caption{Results of squared sound velocity for three-color QCD with the finite isospin chemical potential, originally obtained in Refs.~\cite{Brandt:2022hwy} (left) and~\cite{Abbott:2024vhj} (right).}\label{fig:cs-isospin}
\end{figure}
%%%%%%%%%%%%%%%%%%%%%%%%%%%%%%%%%%%%%%%%%%%%%%%%%%%%%%%%%%
In the case of three-color QCD with the finite isospin chemical potential, the left panel of Figure~\ref{fig:cs-isospin} presents results obtained using a similar calculation strategy with QC$_2$D. Thus, a pionic source term is introduced into the action to stabilize the pion condensate in the superfluid phase~\cite{Brandt:2022hwy}, and then thermodynamic quantities such as $p$ and $e$ were computed directly from lattice observables. The speed of sound was subsequently extracted via the relation $c_{\rm s}^2/c^2 = \partial p / \partial e$.
On the other hand, the right panel of Figure~\ref{fig:cs-isospin} shows results from a novel method that does not rely on introducing a chemical potential to the lattice action~\cite{Abbott:2024vhj}. Instead, the calculation is based on measuring correlation functions involving thousands of pions at $\mu=0$, and using those results to infer the grand‐canonical equation of state (i.e. to reconstruct thermodynamic observables at nonzero $\mu$).
Although this approach does not yet provide definitive evidence for a superfluid phase transition, it allows for controlled access to higher isospin-density regimes.

One particularly interesting aspect is the behavior of the speed of sound at the high-density limit. According to the perturbative QCD calculations, it is generally expected that $c_{\rm s}^2/c^2$ should approach the conformal limit of $1/3$ from below (see, for instance, Appendix of Ref.~\cite{Kojo:2021wax}), but some analytical works suggest that it may instead approach this value from above (see e.g. Refs.~\cite{Braun:2022jme, Fujimoto:2020tjc}). It is also worth noting recent studies that address, in a unified framework, the differences between dense QC$_2$D and three-color QCD with isospin chemical potential~\cite{Fujimoto:2024pcd, Fukushima:2024gmp}.
This raises questions about the asymptotic behavior of strongly interacting matter in the high-density regime and calls for further investigation using complementary methods.

In summary, until recently, numerous lattice studies at finite temperatures and small chemical potential had been performed, but no first-principles calculation in QCD or QCD-like theories had shown any violation of the conformal bound. The observation of such a violation at low temperatures and high densities has only emerged in the past few years, marking a significant shift in our understanding of QCD matter. These findings have opened up new directions in the study of dense, strongly interacting matter, motivating further investigations using different lattice formulations and theoretical approaches.

%%%%%%%%%%%%%%%%%%%
\subsection{Related works: Breaking of the conformal bound and implications for neutron star matter}
%%%%%%%%%%%%%%%%%%%
An important open question is whether the violation of the conformal bound observed in QCD-like theories at high density also occurs in real QCD at finite baryon chemical potential. While first-principles lattice simulations at finite baryon density remain hindered by the sign problem, phenomenological and observational studies of neutron stars have provided indirect yet compelling indications.

In the context of neutron star physics, the possibility of a large sound velocity in cold and dense matter has been discussed for over a decade. In particular, early works based on a smooth crossover between hadronic and quark matter phases~\cite{Masuda:2012ed,Baym:2017whm} proposed that the speed of sound squared could exhibit a peak in medium from the data analysis of the mass of neutron stars.
More recently, updated analyses incorporating multi-messenger observational data, such as mass-radius measurements, support this scenario. Several studies report that $c_{\rm s}^2/c^2$ may exceed the conformal limit of $1/3$ in the dense core of neutron stars~\cite{Marczenko:2020jma, Altiparmak:2022bke, Brandes:2022nxa, Annala:2023cwx, Marczenko:2022jhl, Brandes:2023hma, Fujimoto:2024cyv}. Interestingly, some of these works also argue against the presence of a first-order phase transition in the density range relevant to neutron star interiors, challenging long-held assumptions about the QCD phase diagram at low temperature~\cite{Brandes:2023hma}.

Parallel to these astrophysical insights, several effective model studies have explored the thermodynamic properties of dense QCD and QCD-like theories. These include approaches based on quarkyonic models~\cite{McLerran:2018hbz, Fujimoto:2023mzy}, NJL-type models~\cite{Kojo:2021ugu,Kojo:2021hqh}, functional renormalization group~\cite{Braun:2022olp, Braun:2022jme}, and holographic constructions~\cite{Hoyos:2016cob, Ecker:2017fyh, Hoyos:2021uff}. A recurring feature in many of these studies is the emergence of a nonmonotonic structure in $c_{\rm s}^2(\mu)$, often exhibiting a peak that exceeds the conformal bound. 
While the exact mechanism responsible for generating such large values of the speed of sound remains unclear, several recent studies have proposed new insights that are consistent with the emergence of this behavior.

%%%%%%%%%%%%%%%%%%%%%%%%%%%%%%%%%%%%%%%%%%
\section{Conclusions and Outlook}\label{sec:summary}
%%%%%%%%%%%%%%%%%%%%%%%%%%%%%%%%%%%%%%%%%%

In this review, we have summarized recent progress in first-principles lattice Monte Carlo studies of the phase structure and EoS in dense QC$_2$D. Remarkably, the phase structure is now becoming well-understood, thanks to the results obtained by four independent lattice collaborations employing various lattice actions. Despite differences in formulations and simulation details, the findings are roughly consistent, indicating the emergence of a coherent and robust picture of dense matter in QC$_2$D.
As for the EoS, one of the most intriguing developments is the observation that the speed of sound can exceed the conformal bound, $c_{\rm s}^2/c^2 = 1/3$, in the cold and high-density regime. This behavior has now been reported by several groups independently, using different methods and formulations, lending strong support to the reliability of this result.
Further advances in algorithmic techniques and computational resources will enable more precise and systematic studies, including approaches to the continuum limit and the chiral limit. These efforts will deepen our understanding of strongly interacting matter at high density.

Moreover, there are also ongoing works of the dense QC$_2$D on the hadron mass spectrum~\cite{Hands:2007uc, Wilhelm:2019fvp, Murakami:2022lmq}, the hadron-hadron interactions~\cite{Murakami:2023ejc}, and several gluonic correlation functions~\cite{Boz:2018crd, Buividovich:2020dks, Buividovich:2021fsa}. Although QC$_2$D is a QCD-like theory and does not precisely correspond to real-world QCD, it offers a valuable opportunity to explore dense baryonic matter from first-principles calculations. The results reviewed here suggest that the dense regime hosts rich and novel physics that is qualitatively different from the vacuum or high-temperature QGP phases.

\acknowledgments
The author would like to thank K.~Iida, K.~Ishiguro,  T.-G.~Lee, K.~Murakami, and D.~Suenaga for their collaboration on the original research underlying this review. We are also deeply grateful to Y.~Fujimoto, S.~Hands, T.~Hatsuda, T.~Kojo, J.-I.~Skullerud and  N.~Yamamoto for consistently providing valuable insights and information. 
The work of E.~I. is supported by 
JSPS Grant-in-Aid for Transformative Research Areas (A) JP21H05190, %ExU
JSPS Grant Number JP23H05439, % Kiban-S
JST Grant Number JPMJPF2221,  % SQAI
JPMJCR24I3,  % CREST
JSPS Grant Number 25K01001, %Kiban-B
and also supported by Program for Promoting Researches on the Supercomputer ``Fugaku'' (Simulation for basic science: from fundamental laws of particles to creation of nuclei) and (Simulation for basic science: approaching the new quantum era), and Joint Institute for Computational Fundamental Science (JICFuS), Grant Number JPMXP1020230411. %Fugaku
This work is supported by Center for Gravitational Physics and Quantum Information (CGPQI) at Yukawa Institute for Theoretical Physics.

% Bibliography

%% [A] Recommended: using JHEP.bst file
\bibliographystyle{JHEP}
\bibliography{2color}

\end{document}